\documentclass[dvipsnames,11pt]{article}

\usepackage[utf8]{inputenc}
\usepackage[T1]{fontenc}
\usepackage[sans]{dsfont}
\usepackage[mathscr]{eucal}
\usepackage{amsmath,amsthm,amsfonts,amssymb,mathtools,bbm,mathdots,tensor,bigints}
\usepackage{authblk}   %
\usepackage{braket}		%
\usepackage{graphicx}
\usepackage{slashed}
\usepackage{tikz-cd}

\usepackage[]{xcolor}
\usepackage[]{hyperref}
\hypersetup{colorlinks = true, linktoc=page, urlcolor=MidnightBlue, citecolor=red, linkcolor=ao!90!black}
\definecolor{ao}{rgb}{0.0, 0.5, 0.0}
\usepackage[hyperpageref]{backref}
\usepackage{tikz}
\usetikzlibrary{decorations.pathreplacing,decorations.markings,arrows.meta}
\tikzset{->-/.style={decoration={
			markings,
			mark=at position .58 with {\arrow{>[scale=2]}}},postaction={decorate}}}
\usepackage{tikz-cd}
\usepackage{nicematrix}
\usepackage[sort,compress]{cite}

\usepackage[top=20mm,left=25mm, right=25mm]{geometry}

\renewcommand{\d}{\mathrm{d}}

\newcommand{\spa}{z}

\usepackage{comment}

\newcommand{\msf}[1]{\mathsf{#1}}
\newcommand{\mfk}[1]{\mathfrak{#1}}
\newcommand{\mbf}[1]{\mathbf{#1}}
\newcommand{\mbb}[1]{\mathbb{#1}}
\newcommand{\mscr}[1]{\mathscr{#1}}
\newcommand{\mcal}[1]{\mathcal{#1}}
\newcommand{\mds}[1]{\mathds{#1}}

\newcommand{\wt}[1]{\widetilde{#1}}
\newcommand{\wh}[1]{\widehat{#1}}

\newcommand{\bw}{\bar{w}}

\newcommand{\A}{\msf{A}}

\newcommand{\CP}{\mathbb{CP}}
\newcommand{\im}{\mathrm{i}}
\newcommand{\C}{\mathbb{C}}
\newcommand{\rd}{\mathrm{d}}
\newcommand{\tr}{\mathrm{tr}}
\newcommand{\SU}{\mathrm{SU}}
\newcommand{\cA}{\mathcal{A}}
\newcommand{\End}{\mathrm{End}}
\newcommand{\dbar}{\bar\partial}
\newcommand{\be}{\begin{equation}}
\newcommand{\ee}{\end{equation}}
\newcommand{\p}{\partial}
\newcommand{\Z}{\mathbb{Z}}
\newcommand{\f}{f}
\newcommand{\R}{\mathbb{R}}
\newcommand{\cD}{\mathcal{D}}
\newcommand{\e}{\mathrm{e}}
\newcommand{\diag}{\, \mathrm{diag}}

\newcommand{\hit}{\mscr{M}_{\text{H}}(\Sigma,G)}

\newcommand*\ovl[1]{%
  \vbox{%
    \hrule height 0.8pt%
    \kern 0.25ex%
    \hbox{%
      \kern 0.0em%
      \ifmmode#1\else\ensuremath{#1}\fi%
      \kern 0.0em%
    }%
  }%
}

\newenvironment{eqaligned}
{%
\begin{equation}
    \begin{aligned}
    } 
{%
\end{aligned}
\end{equation}
\ignorespacesafterend}

\newenvironment{eqaligned*}
{%
\begin{equation*}
    \begin{aligned}
    } 
{%
\end{aligned}
\end{equation*}
\ignorespacesafterend}

\newenvironment{eqgathered}
{%
\begin{equation}
    \begin{gathered}
    } 
{%
\end{gathered}
\end{equation}
\ignorespacesafterend}

\newenvironment{eqgathered*}
{%
\begin{equation*}
    \begin{gathered}
    } 
{%
\end{gathered}
\end{equation*}
\ignorespacesafterend}

\theoremstyle{definition}

\theoremstyle{definition}
\newtheorem{lem}[]{Lemma}[section]

\theoremstyle{definition}

\theoremstyle{definition}
\newtheorem{prop}{Proposition}[section]

\theoremstyle{definition}
\newtheorem*{prop*}{Proposition}

\theoremstyle{definition}
\newtheorem{rmk}[]{Remark}[section]

\theoremstyle{definition}

\theoremstyle{definition}
\newtheorem*{prf}{Proof}

\allowdisplaybreaks
\numberwithin{equation}{section}

\usepackage{pgfplots}
\pgfplotsset{compat=1.18}

\makeatletter
\renewcommand{\paragraph}{%
  \@startsection{paragraph}{4}%
  {\z@}{2.25ex \@plus 1ex \@minus .2ex}{-1em}%
  {\fontsize{10.4}{10}\selectfont\bfseries}%
}
\makeatother

\renewcommand{\footnoterule}{\vfill\kern -3pt \hrule width 0.4\columnwidth \kern 2.6pt}

\newlength{\bibitemsep}
\newlength{\bibparskip}
\let\oldthebibliography\thebibliography
\renewcommand\thebibliography[1]{%
  \oldthebibliography{#1}%
  \setlength{\parskip}{\bibitemsep}%
  \setlength{\itemsep}{\bibparskip}%
}

\title{\bf\LARGE\centering The Self-Duality Equations on a Riemann Surface and Four-Dimensional Chern--Simons Theory}

\author[a]{Roland Bittleston\footnote{\href{mailto:rbittleston@perimeterinstitute.ca}{\texttt{rbittleston@perimeterinstitute.ca}}}}
\author[b]{Lionel Mason\footnote{\href{mailto:lmason@maths.ox.ac.uk}{\texttt{lmason@maths.ox.ac.uk}}}}
\author[b,c]{Seyed Faroogh Moosavian\footnote{\href{mailto:sfmoosavian@gmail.com}{\texttt{sfmoosavian@gmail.com}}}}

\affil[a]{\small Perimeter Institute for Theoretical Physics, 31 Caroline Street, Waterloo,
ON N2L 2Y5, Canada}
\affil[b]{\small Mathematical Institute, University of Oxford, Woodstock Road, Oxford, OX2 6GG, United Kingdom}
\affil[c]{Shanghai Institute for Mathematics and Interdisciplinary Sciences, Block A, International Innovation Plaza, No. 657 Songhu Road, Yangpu District, Shanghai, China}

\date{}

\begin{document}

\maketitle

\begin{abstract}
We construct a Lagrangian formulation of Hitchin’s self-duality equations on a Riemann surface $\Sigma$ using potentials for the connection and Higgs field. This two-dimensional action is then obtained from a  four-dimensional Chern--Simons theory on $\Sigma\times \CP^1$  with  an appropriate choice of meromorphic 1-form on $\CP^1$ and boundary conditions at its poles. We show that the symplectic structure induced by the four-dimensional theory coincides with the canonical symplectic form on the Hitchin moduli space in the complex structure corresponding to the moduli space of Higgs bundles. We further provide a direct construction of Hitchin Hamiltonians in terms of the four-dimensional gauge field. Exploiting the freedom in the choice of the meromorphic one-form, we construct a family of four-dimensional Chern--Simons theories depending on a $\mathbb{CP}^1$-valued parameter. Upon reduction to two dimensions, these descend to a corresponding family of two-dimensional actions on $\Sigma$ whose field equations are again Hitchin’s equations. Furthermore, we obtain a family of symplectic structures from our family of four-dimensional theories and show that they agree with the hyperk\"ahler family of symplectic forms on the Hitchin moduli space, thereby identifying the $\mathbb{CP}^1$-valued parameter with the twistor parameter of the Hitchin moduli space.  Our results place Hitchin’s equations and their integrable structure within the framework of four-dimensional Chern--Simons theory and make the role of the twistor parameter manifest.

\end{abstract}

\pagenumbering{gobble}
\pagebreak 
\pagenumbering{roman}
\tableofcontents
\pagenumbering{arabic}

\section{Introduction}
\label{sec:introduction}

Integrable systems play a central role in mathematics and physics, linking diverse areas of these disciplines. There have been various attempts to classify such systems systematically: unifying their constructions and organizing them into natural families. This paper fills in omissions in two such approaches to integrable systems.  The first approach is based on reductions of the self-duality equations and their twistor correspondence \cite{MasonWoodhouse199605,Costello202004,BittlestonSkinner202011}, and the second is based on 4d Chern--Simons theory \cite{Costello201303,CostelloYamazaki201908}, which provides a Lagrangian formulation of integrability primarily for systems in two dimensions. 
We focus on a particularly important family of examples in two dimensions that includes the  Hitchin self-duality equations on a Riemann surface \cite{Hitchin198707}.  These were first obtained by direct reduction from the self-dual Yang--Mills equations in four dimensions. We explain how they and some of their associated structures arise in 4d Chern--Simons theory. 

\subsection{General Themes}

 In the monograph \cite{MasonWoodhouse199605}, a programme was set out to develop an overview of the theory of integrable systems by expressing them as symmetry reductions of the self-duality equations in four-dimensions following earlier observations by Ward \cite{Ward:1985gz}.  The underlying aim was to derive the theory of such integrable equations by reduction of the associated twistor correspondences between solutions to the self-duality equations in four-dimensions and complex analytic objects on twistor space;  for self-dual gravity this was the nonlinear graviton construction of Penrose \cite{Penrose197601}, and for the self-dual Yang--Mills theory, it is the Ward construction \cite{Ward:1977ta}.  Although \cite{MasonWoodhouse199605} gave a good classification of a large class of integrable systems and reformulations of many of their integrability structures into twistor space, there were at least two major gaps in the story.  The first was in the treatment of Hamiltonian, symplectic, and Lagrangian structures;  these were essentially treated by reverse engineering from space-time formulae.  Such structures are foundational for many aspects of integrability, from understanding their hierarchies of conserved quantities and associated commuting flows, to the formulation of R-matrices and quantization.  A second major gap was in the twistor treatment of the two-translation reductions of the self-duality equations.  This is because the quotient by the action of the complexified symmetry groups is singular.  This can be contrasted with the case of a commuting translation and rotation for which the quotient is a Riemann surface, and, although non-Hausdorff, leads to an efficient twistor construction \cite{Woodhouse:1988}.

\smallskip More recently, 4d Chern--Simons (CS) theory has been introduced as a vehicle to understand integrability in two dimensions \cite{Costello201303}. The theory is defined on a four-manifold of the form $\Sigma \times C$, 
and depends holomorphically on the coordinate on $C$ while depending only topologically on the coordinate on $\Sigma$; accordingly, $C$ and $\Sigma$ are referred to as the holomorphic and topological planes, respectively.\footnote{ However, boundary conditions will later introduce a complex structure on $\Sigma$.} Its action functional is given by\footnote{We have chosen the overall normalization of the action in such a way that certain formulas, in particular the symplectic structures, match the standard conventions in the literature.}
\begin{equation}\label{eq:action 4d CS, introduction}
    S_{\text{CS}_4}[\A]= \frac{\im}{8\pi^2}\bigintsss_{\Sigma\times C}\omega\wedge\text{CS}(\A).
\end{equation}
Here $\omega$ is a given meromorphic $(1,0)$-form on $C$, and $\A$, which is a connection on a vector bundle $E\to\Sigma\times C$, is a 1-form with values in the complexified Lie algebra $\mathfrak{g}_\C$, the gauge Lie algebra of the theory, and
\begin{equation}\label{eq:CS three-form, introduction}
    \text{CS}(\A):=\tr\left(\A\wedge\rd\A+\frac{2}{3}\A\wedge\A\wedge\A\right),
\end{equation}
is the usual Chern--Simons three-form.
The theory  \eqref{eq:action 4d CS, introduction} depends on the choice of the holomorphic volume form $\omega$ on $C$ which in general must have poles and possibly also zeroes. These zeroes and poles can be interpreted as defects and must be supplemented by boundary conditions. Once these are specified, they give a route to a Lagrangian formulation that incorporates the Lax pair of an integrable system into its Lagrangian formulation, with the integrability of the Lax pair being the primary field equation arising from the four-dimensional Chern--Simons theory. Initial applications were to harmonic maps or sigma models, and Wess--Zumino--Witten\footnote{These theories are sometimes also referred to as  Wess--Zumino--Novikov--Witten models.} (WZW) models  \cite{CostelloYamazaki201908}.  It has gone on to play an important role in unifying a large class of both discrete \cite{CostelloYamazakiWitten201809,CostelloYamazakiWitten201802} and continuous \cite{CostelloYamazaki201908} integrable models in $1+1$ dimensions, with many more applications and developments such as   \cite{IshtiaqueMoosavianZhou201809,CostelloYagi201810,BittlestonSkinner201903,Vicedo201908,DelducLacroixMagroVicedo201909,AshwinkumarTan201910,Schmidtt201912,FukushimaSakamotoYoshida202003,CostelloStefanski202005,Tian202005,TianHeChen202007,BeniniSchenkelVicedo202008,Stedman202009,BittlestonSkinner202011,CaudrelierStoppatoVicedo202012,Zucchini202101,FukushimaSakamotoYoshida202105,Schmidtt202109,IshtiaqueMoosavianRaghavendranYagi202110,FukushimaSakamotoYoshida202112,BoujakhroutSaidi202204,BoujakhroutSaidi202207,IshtiaqueZhou202211,LiniadoVicedo202301,BoujakhroutSaidiLaamaraDrissi202303,Khan202209,HeTianChen202105,Levine202309,Schmidtt202307,AshwinkumarSakamotoYamazaki202309,MoosavianYamazakiZhou202502,LacroixLevineWallberg202505,Schmidtt202508,SakamotoTateoYamazaki202509}.

\smallskip Motivated by all the above, holomorphic Chern--Simons theories were introduced in the full 6d twistor space \cite{Costello202004,BittlestonSkinner202011}.  These give a Lagrangian underpinning to the original Ward construction \cite{Ward:1977ta} adapted to give Lagrangian descriptions of the self-dual Yang--Mills equations in 4d in a gauge adapted to its realization as a 4d WZW model \cite{LosevMooreNekrasovShatashvili199509}. Although holomorphic Chern--Simons formulations of the self-dual sector of maximally supersymmetric Yang--Mills have been around for some time \cite{Witten:2003nn,Boels:2006ir}  (or 
holomorphic BF  theories in the non-supersymmetric case \cite{Mason:2005zm}) these incorporate the antiselfdual (ASD) field also, rather than treating self-dual Yang--Mills as a self-contained theory. Because twistor space, $\CP^3$, is not Calabi--Yau, it has no non-singular holomorphic  $3$-form so the theories of \cite{Costello202004,BittlestonSkinner202011} used  a holomorphic 3-form  with poles. Like 4d Chern--Simons theory, for consistency, these theories need boundary conditions at the poles.  The corresponding symmetry reduction of these 6d Chern--Simons theories by the two real symmetries then directly yields the 4d Chern--Simons treatment of the reductions.  The problem of fixed points and consequent singularities in the reduced twistor spaces mentioned above only arises if one attempts to reduce the twistor spaces as complex manifolds using the imaginary parts of the holomorphic symmetries.  Thus, 4d Chern--Simons theory is only partially complex but nevertheless captures the essentials of non-singular reductions of Ward's twistor construction for these examples in a Dolbeault framework.

\smallskip A key motivation for this paper is to complete the realization of integrable models via four-dimensional Chern--Simons theory to the full range of 2d integrable systems obtained as reductions on a nondegenerate 2-plane given in \cite[\S 6.2]{MasonWoodhouse199605}. A particularly important example is that of Hitchin's self-duality equations on a Riemann surface \cite{Hitchin198707}, whose realization in four-dimensional Chern--Simons theory is the main subject of the present work. They now play multiple fundamental roles across mathematics and physics  \cite{BeauvilleNarasimhanRamanan198907,BeilinsonDrinfeld1991,BeilinsonDrinfeld1996,HauselThaddeus200205,Ngo200406,Ngo200801,DonagiPantev200604,BershadskyJohansenSadovVafa199501,KapustinWitten200604,FrenkelWitten200710,DonagiWitten199510,GaiottoMooreNeitzke200907,NekrasovWitten201002,Gaiotto200911,NekrasovPestun201211}.
Hitchin's equations were first realized as the symmetry reduction of self-dual Yang--Mills equations in four dimensions of Euclidean signature, with two translation symmetries \cite{Hitchin198707} and a non-degenerate quotient.  Remarkably, on reduction to two dimensions, the equations acquire two-dimensional conformal invariance and, as such, depend only on the complex structure of the surface $\Sigma$ on which they are defined. These equations are given by
\begin{eqgathered}\label{eq:Hitchin's equations, introduction}
    F_A + \phi\wedge\phi=0\,,
    \\
    D_A\phi=0, \qquad D_A\star\phi=0,
\end{eqgathered}
where $F_A = \rd A+A\wedge A$ is the curvature of connection $A$ on a vector bundle $E\to\Sigma$, and $\phi$ is a one-form with values in $\text{End}(E)$, $D_A:=\rd+[A,\cdot\,]$ is the covariant derivative, and $\star$ is the Hodge star operation on $\Sigma$ (so the equations   depend on the complex structure of $\Sigma$).
Other examples in this class of non-degenerate 2-translation reductions of self-dual Yang--Mills theory include also harmonic maps to Lie groups or homogeneous spaces (i.e., the principal chiral models) and the WZW  models. In \cite{MasonWoodhouse199605}, these reductions were distinguished by choice of reality condition, gauge choices, and potential formulation of the theories.  In 4d Chern--Simons theory, the various reductions will arise from different choices of holomorphic 1-form $\omega$ and boundary conditions. These were the original focus of \cite{CostelloYamazaki201908}, so we do not discuss their 4d Chern--Simons formulation in detail here.  Further examples include the Toda field theories, which are most easily understood as reductions of the Hitchin equations and will be treated in \S\ref{sec:affineToda}.

\smallskip
Another motivation in this work is to provide a Lagrangian understanding of canonical structures on the moduli spaces of solutions.  In particular, Lagrangians lead to symplectic structures and Hamiltonians on the space of solutions of the corresponding equations, i.e., their moduli space.  A particularly important case is when $\Sigma$ is taken to be a genus-$g\geq 2$ compact Riemann surface without boundary. Hitchin shows that the moduli space $\hit$ of solutions to \eqref{eq:Hitchin's equations, introduction} on $\Sigma$ has a hyperk\"ahler structure, so that it has a metric with three distinguished metric-compatible, covariantly-constant complex structures denoted by $\mathcal{I},\mathcal{J},$ and $\mathcal{K}$ that satisfy the quaternion relations  \cite{Hitchin198707}.  Lowering the index with the metric gives three corresponding symplectic forms denoted by $\Omega_{\mathcal{I}},\Omega_{\mathcal{J}},$ and $\Omega_{\mathcal{K}}$, given in \eqref{eq:symplectic structure in the complex structure I} and \eqref{eq:symplectic structures in the complex structures J and K}.  These complex structures and symplectic forms naturally combine into a $\mbb{CP}^1$-family of complex structures $\mathcal{J}_\zeta$ and corresponding symplectic forms parametrized by $\zeta\in\mbb{CP}^1$ given explicitly by
\begin{equation}\label{eq:family of symplectic structure on Hitchin moduli space}
    \Omega_{\mathcal{J}_\zeta}:=\mbf{a} %
    \, \Omega_\mathcal{I}+\mbf{b} %
    \,\Omega_\mathcal{J}+\mbf{c} %
    \,\Omega_\mathcal{K},
\end{equation}
where
\begin{equation}\label{eq:coefficients a,b,c of J_zeta}
    \left(\mbf{a},\mbf{b},\mbf{c}\right) :=\frac{1}{1+|\zeta|^2} \left(1-|\zeta|^2,\im(\zeta-\bar{\zeta}),(\zeta+\bar{\zeta})\right)\, .
\end{equation}
According to the choice of $\zeta$, $\hit$ has two different interpretations. In the complex structures $\pm\,\mathcal{I}$ corresponding to $\zeta=0,\infty$, it is the moduli space of Higgs bundles\footnote{A Higgs bundle is a pair $(E,\varphi)$ where $E$ is a holomorphic vector bundle $E\to\Sigma$, and $\varphi\in \Omega^{1,0}_\Sigma(\text{End}(E))$ is an $\text{End}(E)$-valued holomorphic $1$-form, the $(1,0)$ part of $\phi$. 
} satisfying a suitable stability condition.
Assuming certain (semi-)stability conditions, $\hit\simeq T^*\mscr{M}_{\text{Hol}}(\Sigma,G)$, where $\mscr{M}_{\text{Hol}}$ denotes the moduli space of (semi-)stable holomorphic bundles on $\Sigma$, and $\simeq$ should be understood as a birational equivalence of the two spaces. 
Hitchin then shows that $\Omega_{\mathcal{J}}+i\Omega_{\mathcal{K}}$  %
defines a holomorphic symplectic form for the complex structure $\mathcal{I}$, and finds a complete set of $\mathcal{I}$-holomorphic commuting Hamiltonians realizing $\hit$ as the phase space of an algebraically completely Hamiltonian integrable system \cite{Hitchin198701}, see   \S\ref{sec:symplectic structure, and Hitchin Hamiltonians} for more details.  The second interpretation follows for all complex structures $\mathcal{J}_\zeta$ for  $\zeta\ne 0,\infty$ where $\hit$ can be identified with the moduli space of complexified flat connections, also known as the character variety.
The relationship between these two interpretations of $\hit$ is the subject of non-Abelian Hodge correspondence developed in \cite{Donaldson198707,Corlette198801,Hitchin199207,Simpson198810,Simpson199212}. 
These constructions, including the definition of Higgs bundles, can be extended to noncompact surfaces with punctures equipped with parabolic structures \cite{Simpson199007,Konno199304,Yokogawa199301,BodenYokogawa,Sabbah199905,BiquardBoalch200401,Boalch199506} (and also higher-dimensional K\"ahler manifolds \cite{Simpson199212}).  The actions of 4d Chern--Simons theory, which lead to $\Omega_{\mathcal{I}}$ and $\Omega_{\mathcal{J}_\zeta}$ will be presented in \S\ref{sec:from 4d CS theory to Hitchin's equations} and \S\ref{sec:CP^1-family of 2d actions associated with Hitchin's equations}, respectively.

\subsection{Contributions of the Paper}

We now summarize the main results of this work. We will throughout take $C=\CP^1$, and $w$ and $z$ to be local holomorphic coordinates on the $\Sigma$ and $\CP^1$ factors, respectively.  For the most part we will take  $G=\SU(n)$ and $\text{Lie}(G)=\mfk{su}(n)$ although extensions to more general $G$ will be straightforward and stated where simple to do so.

\paragraph{2d Action for Hitchin's equations (\S\ref{sec:2d action for Hitchin's equations}).} We begin in \S\ref{sec:2d action for Hitchin's equations} by introducing the following 2d action on $\Sigma$
\begin{eqaligned} \label{eq:action for Hitchin's equations, introduction}
    S_{\mathrm{H}} [h,\psi,\wt{\psi}]=S_{\mathrm{WZW}}[h]  + \bigintsss_\Sigma  |\partial \psi|^2_h  \,.
\end{eqaligned}
Here,  $h:\Sigma\to \text{Herm}_0$,  $\psi:\Sigma\to\text{Lie}(G)$, where $|\cdot|_h$ is defined in \eqref{eq:action for Hitchin equations} and  $S_{\mathrm{WZW}}[h]$ is the Wess--Zumino--Witten action, defined in \eqref{eq:WZW action for h}.
For $G=\text{SU}(n)$,  
$h\in\mathrm{Herm}_0$, the space of positive definite Hermitian matrices of unit determinant, 
since we can write $h = f_0^*f_0$ for $f_0:\Sigma\to\mathrm{SL}(n,\C)$ identifying the coset  $\mathrm{SU}(n)\setminus\mathrm{SL}(n,\C)$ with $h\in\mathrm{Herm}_0$.
   Using $\rd_\Sigma=\partial+\bar{\partial}$ and the identification
\begin{equation}
    A=h^{-1}\partial h, \qquad \phi=h^{-1}\partial\psi h+\bar\partial\wt{\psi}, 
\end{equation}
we show that the field equations and gauge freedoms of \eqref{eq:action for Hitchin's equations, introduction} coincide with Hitchin's equations \eqref{eq:Hitchin's equations, introduction} in a holomorphic gauge.

\paragraph{2d Action for Hitchin's equations from 4d CS Theory (\S\ref{sec:4dCS-Hitchin}--\ref{sec:symplectic structure, and Hitchin Hamiltonians}).} After presenting our four-dimensional Chern--Simons theory framework in \S\ref{sec:4d CS theory setup}, we derive the action \eqref{eq:action for Hitchin's equations, introduction} from 4d Chern--Simons theory. More specifically, we prove the following result in \S\ref{sec:4dCS-Hitchin} and \S\ref{sec:symplectic structure, and Hitchin Hamiltonians}.
\begin{prop}\label{prop:main proposition, complex structure I}
Take the four-dimensional Chern--Simons theory \eqref{eq:action 4d CS, introduction} with 
\begin{equation}
 \omega=\frac{\rd\spa}{\spa},  \qquad G_\C=\text{SL}(n,\C)\, .
\end{equation}
Since  $\omega$ has poles at $z=0,\infty$, we impose   the  boundary conditions
\begin{equation}
    \label{A-bc0}
\left(\A_w,\A_{\bar w}\right)   \sim 
\left\{
\begin{aligned}
    &\left(\spa^{-1}\,, \spa^2\right) &\qquad\text{as}\qquad \spa&\to0\,,
    \\
    &\left(\spa^{\,-2},\spa\right) &\qquad \text{as}\qquad \spa&\rightarrow \infty\, .
\end{aligned}\right.
\end{equation}
Then, if the field equations are imposed, there  exists  a gauge in which the new gauge field $\cA$ can be written as
\begin{equation}\label{eq:new gauge field cA, introduction}
    \cA = \frac{1}{\spa}\varphi+A+\spa\wt\varphi\,,
\end{equation}
and the action \eqref{eq:action 4d CS, introduction} is gauge equivalent to \eqref{eq:action for Hitchin's equations, introduction} for the complexified Hitchin equations, defined by the pair $(A,\phi)$ in \eqref{eq:new gauge field cA, introduction} with $\phi=\varphi+\wt\varphi$, locally in $\Sigma$. If we further impose the reality condition 
\begin{equation}\label{eq:reality condition on 4d gauge field, introduction}
     \A(-1/\bar\spa)= - \A^*(\spa)\,,
\end{equation}
then we arrive at the Hitchin equations with standard reality conditions.  Here $\A(-1/\bar\spa)$ denotes the pullback of $\A$ by the map $z\mapsto -1/\bar\spa$, which exchanges the $(0,1)$-form part of $\A$ in the $\CP^1$ direction for a $(1,0)$-form.  This is compatible with equation \eqref{eq:reality condition on 4d gauge field, introduction}, since Hermitian conjugation also exchanges $(0,1)$- and $(1,0)$-forms. 

\smallskip We go on to derive the symplectic structure $\Omega_{\mcal{I}}$ given in \eqref{eq:symplectic structure in the complex structure I} on the space of solutions of the field equations of \eqref{eq:action for Hitchin's equations, introduction} from 4d Chern--Simons theory.  This is the canonical symplectic structure on $\hit$ associated with the complex structure $\mathcal{I}$ appropriate to the identification of $\hit$  with the moduli of suitable Higgs bundles on compact Riemann surfaces. \qed
\end{prop} 

 As a by-product, we can construct Hitchin Hamiltonians directly in terms of the gauge fields of our four-dimensional Chern--Simons theory, as we will explain in \S\ref{sec:symplectic structure, and Hitchin Hamiltonians}.

\paragraph{$\CP^1$-Family of 2d Actions from 4d CS Theory (\S\ref{sec:CP^1-family of 2d actions associated with Hitchin's equations}).} Since  $\hit$ as a hyperk\"ahler manifold has a whole $\CP^1$  family of symplectic structures \eqref{eq:family of symplectic structure on Hitchin moduli space},  we generalize Proposition \ref{prop:main proposition, complex structure I} in \S\ref{sec:CP^1 family of 4d CS theories and the corresponding 2d actions} and \S\ref{sec:hyperkahler family of symplectic structures} to give:

\begin{prop}\label{prop:main proposition, complex structure J_{zeta0}}
    Consider the $\zeta \in\CP^1$ family of 4d Chern--Simons theories \eqref{eq:action 4d CS, introduction} with  
\begin{equation}
 \omega(\zeta ,\bar{\zeta})=\frac{(\zeta +1/ \bar\zeta )^2\spa\rd \spa}{(\spa-\zeta )^2 (\spa +1/\bar\zeta )^2}, %
\end{equation}
so that $\omega(\zeta ,\bar{\zeta})$ has zeroes at $\spa=0,\infty$ where we allow the following boundary condition
\begin{equation}
    \begin{alignedat}{3}
    \A_w&\sim\mathcal{O}(\spa^{-1})\,, &\qquad \spa&\to 0\,,
    \\
    \A_{\bw}&\sim\mathcal{O}(\spa)\,, &\qquad \spa&\to\infty\,.
\end{alignedat}
\end{equation}
At the double poles of $\omega$ at $z=\zeta ,-1/\bar{\zeta}$, we impose the boundary conditions
\begin{equation}
    \begin{alignedat}{3}
        \A_{\bw}&\sim\mathcal{O}((\spa-\zeta)^2)\,,&\qquad &\spa\to\zeta\,,
        \\
        \A_{w}&\sim\mathcal{O}((\spa+1/\bar\zeta)^2)\,, &\qquad &\spa\to - 1/\bar{\zeta}\,.
    \end{alignedat}
\end{equation}
We furthermore impose the reality conditions  \eqref{eq:reality condition on 4d gauge field, introduction}.
Then, we can reduce to a corresponding family of 2d actions $S_{\text{H},\zeta}$, given in \eqref{eq:action for the complex structure J_{zeta_0}}, parametrized by $\CP^1$, which are gauge equivalent to  \eqref{eq:action 4d CS, introduction} for various choices of $\zeta \in\CP^1$. The field equations of this action are then equivalent to Hitchin's equations. Finally, the symplectic structure on the space of solutions of the field equations of $S_{\text{H},\zeta}$ arising from the four-dimensional Chern--Simons theory is $\Omega_{\mcal{J}_{\zeta }}$, given in \eqref{eq:family of symplectic structure on Hitchin moduli space}.
\qed
\end{prop}

This identifies the parameter $\zeta $ of \eqref{eq:family of symplectic structure on Hitchin moduli space} with the twistor parameter of the Hitchin moduli space.
The proof of Proposition \ref{prop:main proposition, complex structure J_{zeta0}} is identical to that of Proposition \ref{prop:main proposition, complex structure I} with the role of $0$ and $\infty$ in the latter played by $\zeta $ and $-1/\bar{\zeta}$ in the former, so we only give a brief discussion. 

\begin{rmk} \label{rmk:other-actions}
    We highlight that there are (at least) two known actions relating to Hitchin's equations:  One is a Lagrange multiplier type action obtained by compactifying 4d mixed topological-holomorphic BF theory to 2d dimensions \cite[\S 7.1]{Jarov202509}.  This is the 2d counterpart of the Chalmers--Siegel \cite{Chalmers:1996rq} action for self-dual Yang--Mills theory, whereas the functional \eqref{eq:action for Hitchin's equations, introduction} is the 2d counterpart of the 4d WZW model, or Donaldson functional.  We also note that the gauged 4d Chern--Simons set-ups of \cite[\S4]{ColeCullinanHoareLiniadoThompson202407} yield 2d integrable field theories coupled to a Hitchin subsector. On the other hand, a variational formulation of the Hitchin integrable system based on a three-dimensional topological-holomorphic version of BF theory has been constructed in \cite[Thms.~4.1 and~4.3]{CaudrelierHarlandSinghVicedo202509} (see also \cite{VicedoWinstone202201,Cole:2025zmq} for related works).  We emphasise that the Hitchin integrable system lives on the moduli space of solutions to Hitchin's equations, so although both functionals have a natural description in terms of mixed topological-holomorphic field theory, they do not appear to be immediately related. \qed 
\end{rmk}

\section{From 4d Chern--Simons Theory to Hitchin's Equations}
\label{sec:from 4d CS theory to Hitchin's equations}

This section first presents an action for Hitchin's equations in \S\ref{sec:2d action for Hitchin's equations} and goes on to define a corresponding four-dimensional Chern--Simons theory as in \eqref{eq:action 4d CS, introduction} in \S\ref{sec:4d CS theory setup}.
Then in \S\ref{sec:4dCS-Hitchin} we give a proof of Proposition \ref{prop:main proposition, complex structure I} that explains how the 2d action can be reduced from that in 4d.  
We show explicitly that in a local patch of $\Sigma$, the action \eqref{eq:action 4d CS, introduction} subject to appropriate boundary conditions at the poles of $\omega$ descends to our 2d Lagrangian that gives rise to the Hitchin equations.   We go on to compute the symplectic structure on the space of solutions of the field equations in \S\ref{sec:symplectic structure, and Hitchin Hamiltonians}, and show that it coincides with the canonical symplectic structure on $\hit$ associated with the complex structure $\mcal{I}$. We go on to explain how to construct the Hitchin Hamiltonians on the moduli space from four-dimensional Chern--Simons theory.

\subsection{2d Action for Hitchin's equations}\label{sec:2d action for Hitchin's equations}

Hitchin's equations \cite{Hitchin198707} arise as the reduction of four-dimensional self-dual Yang--Mills equations in the Euclidean signature to two dimensions.  
Upon reduction, they acquire conformal invariance so that they can be naturally defined on any Riemann surface $\Sigma$. They depend on a choice of structure group which, for simplicity and concreteness, we will take to be $\SU(n)$.
 
\smallskip In the first instance, they are a system of equations on a choice of unitary connections $D=\rd+A$ on a rank-$n$ bundle $E\rightarrow \Sigma$, and a Higgs field $\varphi\in \Omega^{1,0}_\Sigma(\End(E))$, and can be written as\footnote{These equations are equivalent to \eqref{eq:Hitchin's equations, introduction} by writing  $\phi = \varphi + \varphi^*$.} 
\begin{eqaligned} \label{eq:Hitchin's equations}
    F_A + [\varphi,\varphi^*] = 0\,,\qquad\dbar_A\varphi = 0\,.
\end{eqaligned}
where $F_A =\rd A+A\wedge A$ is the curvature of $D$ and $\wt\varphi \in \Omega^{0,1}_\Sigma(\End(E))$.  More generally, we can complexify the system, so that the connection $A$ is complex and $\varphi^*$ replaced by an independent $\wt\varphi$, with reality subsequently imposed by the condition that there exists a unitary gauge in which $A=-A^*$ and $\wt\varphi   =\varphi^*$,  the Hermitian conjugate to $\varphi$. These equations form a 2d integrable system with the Lax pair defined by the connection
\begin{equation} \label{eq:Lax-Hitchin}
    \rd + \mcal{A}(\spa) \coloneqq \rd + \frac{1}{\spa}\varphi + A + \spa \wt \varphi\,.
\end{equation} 
This depends on the auxiliary spectral parameter $\spa\in \CP^1$ and is flat if and only if the Hitchin equations are satisfied.  The solution is real if in a unitary gauge we have
\begin{equation} \label{eq:reality condition on Hitchin's Lax form}
    \cA(-1/\bar\spa) = - \cA(\spa)^*\,.
\end{equation}

In order to write a second-order action of standard type, we introduce local potentials for our fields  $(A,\varphi,\varphi^*)$.  The connection defines a holomorphic structure on $E$ as there is no integrability condition on a Riemann surface.  Then we can find a   local holomorphic frame, but this will not, in general, be unitary, and  the Hermitian metric  will take the general form
\be
h=h(w,\bar w) \in C^\infty_\Sigma(E^*\otimes \bar E^*),  \qquad \det h=1\, ,
\ee
where reality will now follow from $h=h^*$. There is then a unique  connection $A$ (the \emph{Chern connection}) that preserves the Hermitian metric $h$ given in this complex frame by
\begin{eqaligned}
    A = h^{-1}\p h\, . \label{a-h}
\end{eqaligned}
In such a gauge, the  curvature is 
\begin{equation}
F_A=\dbar (h^{-1}\p h) \,.
\end{equation}
Changing the choice of holomorphic frame by $g=g(w)\in \text{SL}(n,\C)$ gives the holomorphic gauge transformations 
$h\rightarrow  g^* h g$.

In this holomorphic gauge we introduce the potentials  $(\psi, \wt \psi) \in C^\infty (\End(\bar E))\oplus C^\infty (\End( E))$ for the Higgs field by
\begin{eqaligned}\label{eq:higgs field in terms of psi}
    (\varphi , \wt\varphi) =(h^{-1}(\p \psi)h,\dbar \wt \psi)
\end{eqaligned}
where, in the first entry, the conjugation by $h$ ensures that both $(\varphi,\wt\varphi)$  take values in $\End(E)$ and the holomorphic gauge \eqref{a-h} is assumed. Alongside the gauge freedom $g(w)$ in the choice of holomorphic frame of $E$, we also have the gauge freedom $\delta (\psi,\wt\psi)= (\chi(\bar w), \wt\chi( w))$ as is clear from \eqref{eq:higgs field in terms of psi}.  Reality then follows when $\wt\psi=\psi^*$. 

\smallskip With these definitions, Hitchin's equations follow from the action
\begin{eqaligned}
    S_{\mathrm{H}}[h,\psi] = S_{\mathrm{WZW}}[h] + \bigintsss_\Sigma |\p \psi|^2_h\,,\qquad |\p\psi|^2_h \coloneqq - 2\im\,\tr (h^{-1}\p\psi h\wedge\dbar \wt\psi)\,.
    \label{eq:action for Hitchin equations}
\end{eqaligned}
Here $S_{\mathrm{WZW}}[h]$ is the Wess--Zumino--Witten action \cite{WessZumino19711,Novikov198210,Witten198308,Witten198412,Witten199102} 
\begin{equation}\label{eq:WZW action for h}
    S_{\mathrm{WZW}}[h]\coloneqq\im\bigintsss_\Sigma\tr\left(h^{-1}\p h \wedge h^{-1} \dbar h\right) + I_{\mathrm{WZ}}[h]\,,
\end{equation}
where
\begin{equation} \label{eq:WZ-term}
    I_\mathrm{WZ}[h] = \frac{\im}{3}\bigintsss_{\Sigma\times [0,1]}\tr\big((h^{-1}\rd h)^3\big)\,.
\end{equation}
In this latter Wess--Zumino term, a factor of $\im$ is included for $h$ Hermitian, and as usual, an extension of $h$ from $\Sigma\times 0$ to $\Sigma\times [0,1]$ has been taken that is fixed to be the identity at $\Sigma\times 1$.  The third homotopy group of the space of unit determinant positive definite Hermitian matrices is trivial,  
($\text{SL}(n,\C)/\SU(n)$ is contractible) 
so there is no ambiguity in the second term, and its normalization does not need to be fixed.\footnote{If we were to change the reality conditions to those for a harmonic map, then we would need to impose the usual normalization that leads to an ambiguity of integral multiples of  $2\pi \im$ in this term.}  This is essentially the two-dimensional version of the action in \cite{Donaldson198501} for Hermitian Yang-Mills due to Donaldson, who already observes its equivalence with the a WZW-type action in \cite{Donaldson:1992vc}, see also \cite{Simpson198810} for applications to Higgs bundles and also \cite[\S 3.4]{MasonWoodhouse199605} for an  alternative local formula.

\smallskip It can be seen directly that the full action yields the Hitchin equations.  The $\wt\psi$ variation of the second term of \eqref{eq:action for Hitchin equations} yields $ \dbar\varphi=0$, as defined in \eqref{eq:higgs field in terms of psi}, which is equivalent to $\dbar_A \varphi=0$ in this holomorphic gauge.  On the other hand, the $h$-variation of $S_{\mathrm{WZW}}$ yields the curvature $F_A$ and the variation of the second term of \eqref{eq:action for Hitchin equations} yields the commutator of the Higgs field with its Hermitian conjugate as required. Furthermore, the action is invariant under the choice of $\psi$ as the second term is the squared norm of the Higgs field. It is also invariant under the choice of a holomorphic frame, as the variation of $h$ just leads to gauge covariant equations of motion.

\subsection{4d CS Theory Setup}\label{sec:4d CS theory setup}

We now explain the version of  4d Chern--Simons theory for the Hitchin equations. Consider the action \eqref{eq:action 4d CS, introduction} on $\Sigma\times \CP^1$, i.e.,  
  $C=\CP^1$;  we will, for the most part, work locally on a Riemann surface $\Sigma$. 
  The gauge field $\A$ of four-dimensional Chern--Simons theory defines a connection $D_\A$ on a bundle $E\rightarrow \Sigma\times \CP^1$ with structure group $G=\mathrm{SL}(n,\C)$. We will assume that $E$ is topologically trivial on the $\CP^1$ factors. Although  $\Sigma$ and $\CP^1$ denote the topological and holomorphic planes of the theory, our boundary conditions will also require a complex structure on $\Sigma$, so we introduce complex coordinates $(w,\spa)$, with $w\in\C$ a local coordinate on $\Sigma$ and $z$ on $\CP^1$. Since $\A$ is  wedged against $\omega$, which is a multiple of $\rd z$, a meromorphic $(1,0)$-form on $\CP^1$, $\A$ is a partial connection lying  $\Omega^1_\Sigma\oplus\Omega^{0,1}_{\CP^1}$. The 1-form $\omega$ must have poles and can also have zeroes, and this data will be supplemented with appropriate boundary conditions on $\A$ at the zeros and poles of $\omega$.

\smallskip The field equations give rise to connections $\A$ whose curvature satisfies 
\begin{equation}\label{eq:equations of motion of 4d CS}
    \omega\wedge F_{\A}=0,
\end{equation}
where $F_{\A}=\rd\A+\A\wedge\A$.  
This gives the Lax pair formulation of integrable systems living on $\Sigma$.  In general, such a Lax pair for an integrable system on $\Sigma$ is a connection that depends on the auxiliary complex parameter  $\spa\in\CP^1$ and whose flatness is equivalent to the original integrable equations.

\smallskip Here we will aim for \eqref{eq:Lax-Hitchin}, the Lax pair  for Hitchin's equations.
 For this, we make the choice 
\begin{equation}\label{eq:omega for the complex structure I}
    \omega=\frac{\rd\spa}{\spa}\,,
\end{equation}
which has simple poles at $\spa=0,\infty$. In general, one needs to choose boundary conditions on $\A$ at the poles of $\omega$.  For a pole at $\spa=0$, the variation of the action gives rise to a boundary contribution 
\begin{equation}
\frac{1}{4\pi}\bigintsss_\Sigma \text{Res}_{\spa=0}\big[\tr  (\A\wedge\delta \A)\big]\,,\label{bc-symplectic-pot}
\end{equation}
and we must impose boundary conditions on the components of $\A$ near $\spa=0$ so that this vanishes for all variations. Thinking of \eqref{bc-symplectic-pot} as a 1-form on the space of $\A$'s, its vanishing for all variations requires that the germ of $\A$ at $\spa =0$ lies in a Lagrangian subspace for the skew form defined by its variation
\begin{equation}
\frac{1}{4\pi}\bigintsss_\Sigma\mathrm{Res}_{z=0}\big[\tr(\delta_1\A\wedge\delta_2 \A)\big]\,. \label{symplectic-pot}
\end{equation}
Given a complex structure on $\Sigma$ with local holomorphic coordinate $w$, there is a natural hierarchy of choices\footnote{If $\omega$ has  a pole of order $q$, at $\spa=0$ the second term becomes $\mcal{O}(\spa^{p+q})$.}
 \begin{equation}
     \A_w \sim \mcal{O}(\spa^{-p})\,,
\qquad \A_{\bar w}  \sim \mcal{O}(\spa^{p+1})\, , \qquad p\in \Z %
\, ,\label{bck}
\end{equation}
and these are also sufficient to make the action bounded and gauge invariant near the pole at $\spa=0$.  With our $\omega$ as \eqref{eq:omega for the complex structure I}, we must make such a choice at both $\spa=0$ and $\infty$.  For the reduction to the WZW model, one takes $p=0$  at  $\spa=0$ and $p=-1 $ at $\spa=\infty$ \cite{CostelloYamazaki201908}.  For the reduction to the action for Hitchin's equations, we impose the  boundary conditions
\begin{equation}
    \label{A-bc}
\left(\A_w,\A_{\bar w}\right)   \sim 
\left\{
\begin{aligned}
    &\left(\spa^{-1}\,, \spa^2\right) &\qquad\text{as}\qquad \spa&\to0\,,
    \\
    &\left(\spa^{\,-2},\spa\right) &\qquad \text{as}\qquad \spa&\rightarrow \infty\, .
\end{aligned}\right.
\end{equation}
Here $\A_w$ and $\A_{\bar w}$ are the projections of $\A$ onto its $(1,0)$- and $(0,1)$-form parts in the $\Sigma$ direction. The notation $\A_w\sim \spa^{-1}$ as $\spa\to0$ indicates that we permit a simple pole in the $(0,1)$-form component of $\A$, without prescribing its residue.
It is clear  that these boundary conditions ensure the vanishing of the sum of boundary terms
\begin{equation}
\frac{1}{4\pi}\bigintsss_\Sigma\Big(\mathrm{Res}_{\spa =0}\big[\tr(\A\wedge\delta\A)\big] - \mathrm{Res}_{\spa =\infty}\big[\tr(\A\wedge\delta\A)\big]\Big)\,, \label{bdy-term}
\end{equation}
arising from the variation of the action
including the contribution from $\spa=\infty$.

\smallskip These choices will lead to the complexified Hitchin's equations. If we impose the reality condition \begin{equation}\label{eq:reality condition on 4d CS gauge field}
     \A(-1/\bar\spa)= - \A^*(\spa)\,.
\end{equation}
then follows from \eqref{eq:reality condition on Hitchin's Lax form} that the corresponding solutions will be real.

\begin{rmk}
  Other natural reductions of the four-dimensional self-duality equations are obtained by different choices of reality condition.  For example, straightforward complex conjugation within a complex Lie group rather than hermitian conjugation will give, for example, $A$ arising as an $\text{so}(n)$ rather than $\text{su}(n)$ gauge connection.  Alternatively, one can lift the conjugation   $z\rightarrow +1/\bar \spa$ that has fixed points so that on the fixed point set $|\spa|=1$, $\A$ as a whole then takes values in the real Lie algebra.  This gives systems that are gauge equivalent to harmonic maps. See \cite[\S 6.2]{MasonWoodhouse199605} for related examples. \qed 
\end{rmk}

\subsection{Derivation of the 2d Action from 4d CS Theory} \label{sec:4dCS-Hitchin}

This section contains our main aim, i.e., the derivation of the action \eqref{eq:action for Hitchin equations} for Hitchin's equations \eqref{eq:Hitchin's equations} from the four-dimensional Chern--Simons theory setup described in \S\ref{sec:4d CS theory setup}.  We will adopt the familiar strategy from \cite{CostelloYamazaki201908}: First, one enters the gauge in which the $\rd\bar\spa$ component of $\A$ vanishes.  In this gauge the components of the equations of motion involving $\rd\bar\spa$ imply that $\A$ is meromorphic and gauge-equivalent to the Lax form \eqref{eq:Lax-Hitchin}.  This is non-trivial as the Lax form \eqref{eq:Lax-Hitchin} does not satisfy \eqref{A-bc}, so the requisite gauge transformation is incompatible with the boundary conditions.

\smallskip The first step is to prove the following:

\begin{lem}
    Working locally in $\Sigma$, for $\A$ sufficiently small (or generic) there is a one to one correspondence between the gauge-equivalent solutions to the equations of motion of $S_{\text{CS}_4}[\A]$, given in \eqref{eq:action 4d CS, introduction} subject to the boundary conditions \eqref{A-bc}, and local complexified solutions to Hitchin's equations modulo their natural gauge freedom.
\end{lem}

\begin{prf}
In the following, all our considerations will be local on $\Sigma$ and we will abuse notation by using $\Sigma$ for a sufficiently small open subset of $\Sigma$.

\smallskip Since we take our bundle to be topologically trivial on the $\CP^1$-factor, it is generically holomorphically trivial also; although the holomorphic type can jump, it does so in complex co-dimension one and so we assume either that $\A$ small enough that it does not do so, or that it does so only at isolated values of $w$ that we exclude in the first instance.\footnote{Positivity conditions can be introduced that prevent such jumping, see for example \cite[Proposition 9.3.6]{MasonWoodhouse199605}.} Thus locally on $\Sigma$ but globally on $\CP^1$ there exists a   frame $\f:\Sigma\times\CP^1\to G_\C$ that is holomorphic on $\CP^1$ so that  
\begin{equation}
    \label{eq:trivialise} \A_{\bar\spa } = \f^{-1}\partial_{\bar\spa }\f\,.
\end{equation}
Given $\A_{\bar\spa}$, $f$ is unique up to left multiplication 
\begin{equation}\label{eq:gauge freedom in f}
\f\rightarrow \gamma\f,\qquad \mbox{ where }\qquad\gamma=\gamma(w,\bar w)\,.
\end{equation}
This is because such a $\gamma$ must be holomorphic in $\spa$ to maintain $\A_{\bar\spa}=0$ and global on $\CP^1$ so by Liouville's theorem it's independent of $\spa$.

\smallskip We now use $\f$ to transform to a gauge in which the $\bar\spa$ component of the new gauge field $\A^\f  $ vanishes, i.e.  
\begin{equation}\label{eq:new gauge field A'}
    \A^\f  =f\A f^{-1} -\rd f f^{-1}\, , \qquad \rd\coloneqq\rd_{\Sigma}+\ovl{\partial}_{\CP^1},
\end{equation}
such that 
\begin{equation}
     \A^\f  _{\bar\spa}=0\,.\label{Hol-gauge}
\end{equation}

\smallskip However, the boundary conditions \eqref{A-bc} restricts the gauge freedom at $\spa=0,\infty$: a complex gauge transformation   $g:\Sigma\times\CP^1\to G_\C $  preserves \eqref{A-bc} for $\A\rightarrow g\A g^{-1} -(\d g)g^{-1}$ iff 
\be 
\p_{\bar w} g\sim \spa ^2\quad\text{as}\quad\spa \to0\,,\qquad \p_w g\sim \spa ^{-2}\quad\text{as}\quad \spa\to\infty\,. \label{gauge-res}\ee
 Thus,  $g$ must be holomorphic in $w$ to second order in $\spa$ at $\spa=0$ and antiholomorphic in $w$ to second order at $\spa=\infty$.  
However, $f$ does not satisfy these conditions,  and so $\A^\f  $ no longer satisfies \eqref{A-bc}.
If we write 
\begin{equation} \label{eq:connection A0}
\A'  \coloneqq \A^\f   - \Big(\frac{1}{\spa}\varphi + A +\spa\wt\varphi\Big)\,,    
\end{equation}
where  $(A, \varphi,\wt \varphi)\in \Omega^1_\Sigma\oplus \Omega^{1,0}_\Sigma\oplus \Omega^{0,1}_\Sigma$ valued in $\text{End}(E)$ are uniquely determined by
\begin{equation}
    \label{A-prime-bc}
\left(\A^\f  _w,\A^\f  _{\bar w}\right)   = 
\left\{
\begin{aligned}
    &\left (0, A_{\bar w}+  \spa\wt \varphi_{\bar w}\right)+\left(\mcal{O}(\spa^{-1})\,,\mcal{O}( \spa^2)\right) &\qquad\text{as}\qquad \spa&\to0\,,
    \\
    &(A_w +\spa^{-1}\varphi_w ,0)+\left(\mcal{O}(\spa^{\,-2}),\mcal{O}(\spa)\right) &\qquad \text{as}\qquad \spa&\rightarrow \infty,
\end{aligned}\right.
\end{equation}
then  $\A'$ does now  satisfy \eqref{A-bc}.

\smallskip We can now identify $(A,\varphi,\wt\varphi)$  with the ingredients that appear in the Hitchin equations \eqref{eq:Hitchin's equations}.
 If we now impose the holomorphic gauge \eqref{Hol-gauge}, i.e., the vanishing of the $\bar\spa$ components of the curvature, we find that $\A' $ is holomorphic and $(\spa \A' _w, \spa^{-1}\A' _{\bar w})$  are both bounded on $\CP^1$ with the zeroes either at $0$ or $\infty$.  Therefore, $\A' $ must vanish by Liouville's theorem. Thus, 
\begin{equation} \label{eq:gauge field cA in the holomorphic gauge}
\A^\f=\mcal{A}\coloneqq \frac{1}{\spa}\varphi + A + \spa\wt\varphi\,,
\end{equation}
can be identified with the complexified Lax pair \eqref{eq:Lax-Hitchin} for the Hitchin system.  
If we now impose the vanishing of the $w\bar w$ component of the equations of motion \eqref{eq:equations of motion of 4d CS} of 4d Chern--Simons theory, we then obtain the flatness of the Lax pair and hence the complexified Hitchin equations on-shell. \qed 
\end{prf}

\paragraph{Reality Conditions.} If we now impose the reality condition \eqref{eq:reality condition on 4d CS gauge field} on $\A$, then,  if $\f$ satisfies \eqref{eq:trivialise}, so must $\f^*(-1/\bar\spa)^{-1}$. The  solutions to \eqref{eq:trivialise}  are unique up to  \eqref{eq:gauge freedom in f}, so we must have
\begin{equation}
    \f^*(-1/\bar\spa)^{-1}=h\f\, , \qquad h=h(w,\bar w)\, .
\end{equation}  
Evaluating this relation at both  $\spa=0,\infty$ we see
\begin{equation}\label{hermitian}
    h=f_0^{-1}\f_\infty^{*-1}=f_\infty^{-1} f_0^{*-1}\, ,
\end{equation}
where we have defined
\begin{equation}\label{eq:definition of f_0 and f_infty}
     f_0\coloneqq f\big|_{\spa=0}\, , \qquad f_\infty\coloneqq f\big|_{\spa=\infty}\, .
 \end{equation}
Thus $h=h^*$, and hence is Hermitian. It follows that, using a $\gamma$-gauge transformation \eqref{eq:gauge freedom in f} on $\f$ we have
\begin{equation}
   h\rightarrow  \gamma^* h\gamma
\end{equation}
and this can be used to reduce $h$ to the identity, assuming that $h$ is positive definite (otherwise to normal form with $\pm 1$ on the diagonal according to signature).   We define this to be a \emph{unitary gauge} in which the residual $\gamma$-gauge freedom is unitary, $\gamma^*\gamma=I$, yielding the residual gauge freedom of the Hitchin equations as normally presented. With this,  $\A^\f  $ in turn satisfies $\A^\f  (-1/\bar\spa)=-\A^{\f*}(\spa)$ so that $A$ is skew-Hermitian and $\wt \varphi=\varphi^*$ as required.

\paragraph{Potentials.} We now  obtain the potentials $(h,\psi,\wt\psi)$ for $(A,\varphi,\wt\varphi)$ in terms of the asymptotics of $f$.  The boundary condition violating terms in \eqref{A-prime-bc} arise in \eqref{eq:new gauge field A'} from the term 
\begin{equation}
    J\coloneqq\rd f f^{-1}\, .\label{J-def}
\end{equation}
In order to identify them,  expand $f$ around $z=0,\infty$ as
\begin{eqaligned}\label{psi-def}
    f&= f_0(1-z \wt \psi) + \mcal{O}(z^2), &\qquad\text{as }\qquad z&\to 0\, ,
    \\ 
f&=f_\infty (1- \psi/z) + \mcal{O}(z^{-2}), &\qquad \text{as }\qquad  z&\rightarrow\infty\, .
\end{eqaligned}
The residual gauge transformations  $g$ of \eqref{gauge-res} can also be expanded as
\begin{eqaligned}\label{eq:expansion of g near 0 and infty}
    g& =g_0(w)(1+z \chi(w))+\ldots, &\qquad \spa &\to0,\\
    g& =g_\infty(\bar w) (1+ \wt \chi(\bar w)/z)+\ldots, &\qquad \spa&\to\infty\, .
\end{eqaligned}
The boundary conditions \eqref{gauge-res} imply that $g_0 ,\chi$ are holomorphic in $w$  and $g_\infty , \wt \chi$ anti-holomorphic in $w$, although otherwise arbitrary. These transform our fields as
\begin{equation} \begin{aligned} \label{psi-gauge}
&(f_0,f_\infty)\rightarrow (f_0g_0(w),f_\infty g_\infty(\bar w))\,, \\
&(\wt\psi,\psi)\rightarrow (g_0^{-1} \wt\psi g_0, g_\infty^{-1}\psi g_\infty)+(\chi(w),\wt\chi(\bar w))\,.
\end{aligned} \end{equation}
Thus $(\p_{\bar w}\wt\psi,\p_w\psi)$ are covariant under \eqref{psi-gauge}.

\smallskip The $\gamma$-gauge \eqref{eq:gauge freedom in f} acts on $f_0, f_\infty$ on the left but  
the  `holomorphic Wilson lines'  \cite{Mason:2010yk} 
\begin{equation}
    h=f^{-1}_\infty f_0\, ,\label{h-def}
\end{equation}
are invariant.  It further follows that if we fix the $\gamma$-gauge by setting $f_0=I$, together with our reality conditions, it follows from  \eqref{hermitian} that $h$ coincides with that introduced in \eqref{hermitian} and so with our reality conditions, $h =h^*$ and so is Hermitian.  However, the   gauge freedom $\A\rightarrow g\A g^{-1} - (\rd g)g^{-1}$ acts by   $\f\mapsto\f g$ so that
\begin{equation}
h\rightarrow g_\infty(\bar w)^{-1} h g_0(w) \,,   \label{h-gauge}
\end{equation}
where $g_0=g_0(w)$ and $g_\infty=g_{\infty}(\bar w)$. This is the natural gauge freedom for the Hermitian metric $h$ expressed in a holomorphic frame leading to the Chern connection \eqref{a-h}
\begin{equation}
    A = h^{-1}\p h\, . \label{a-h2}
\end{equation} 
We now use the $\gamma$ gauge freedom \eqref{eq:gauge freedom in f} to go to a \emph{holomorphic gauge} in which we 
fix $f_0=1$.  It is straightforward then to check that 
\begin{equation}
    J'\coloneqq J+ z \bar\p\wt \psi+A +  \frac{1}{z}h^{-1}(\p \psi ) h\,,
\end{equation}
now satisfies \eqref{A-bc}. If we set 
\begin{equation} \label{J-0-def}
(\varphi,\wt\varphi) \coloneqq \big(h^{-1}(\p\psi)h, \bar\p\wt\psi\big)\,,
\end{equation}
then we see that $J'=J+\mathcal{A} $ with $\cA$ as defined in \eqref{eq:gauge field cA in the holomorphic gauge} cancelling the deviation of $J$ from \eqref{A-bc}.

\paragraph{Reduction of the Action.} We now consider the reduction of the action of 4d CS theory to the 2d action  \eqref{eq:action for Hitchin equations} on $\Sigma$ for Hitchin's equations. 
More precisely,

\begin{lem}\label{lem:2d action from 4d cs action}
     The 4d Chern--Simons action $S_{\text{CS}_4}[\A]$ of \eqref{eq:action 4d CS, introduction} reduces to the potential action \eqref{eq:action for Hitchin equations} for the Hitchin equations in the holomorphic gauge \eqref{Hol-gauge}.
\end{lem}
\begin{prf}
    To determine the resulting 2d integrable model, we follow the procedure in \cite{CostelloYamazaki201908}.
The strategy is to gauge fix to the holomorphic gauge \eqref{Hol-gauge}, and integrate out the $\A' $ field, defined in \eqref{eq:connection A0}, to land on  \eqref{eq:action for Hitchin equations} (otherwise said, $\A'$ will be seen to decouple and to vanish on shell).  This is complicated by the fact that the gauge transformation $f$ in \eqref{eq:new gauge field A'} does not satisfy \eqref{gauge-res} so that it generates boundary terms that we now identify.  Under the  local gauge transformation $f$, with $J$ as in \eqref{J-def}, the Chern--Simons form \eqref{eq:CS three-form, introduction} changes as 
\begin{equation} \label{eq:CS-transform}
    \mathrm{CS}(\A) = \mathrm{CS}(\A^\f  ) - \d\, \tr(J \wedge\A^\f  ) - \frac{1}{3}\tr(J^3)\,.
\end{equation}
Recall that our boundary conditions guarantee that the left-hand side vanishes at $\spa=0,\infty$ so that the action \eqref{eq:action 4d CS, introduction} is smooth.  However, the individual terms coming from the wedge product with $\omega$ in the right-hand side will have poles, and so the singular terms must cancel. We can therefore consider the limit as $\epsilon\to 0$ of the integral over a domain $U_\epsilon$ in the $\CP^1$  that excludes discs of radius $\epsilon$ around $\spa=0,\infty$. We break up the action into
\begin{eqaligned} \label{eq:action of 4d CS in terms of gauge-transformed CS form}
S_{\text{CS}_4}[\A] = \frac{\im}{8\pi^2}\lim_{\epsilon\to 0} \bigintsss_{\Sigma\times U_\epsilon} \omega \wedge\Big(\text{CS}(\A^\f  ) - \d\tr(J\wedge\A^\f  ) - \frac{1}{3}\tr(J^3)\Big) \eqqcolon S_0+S_1+S_2\,,
\end{eqaligned}
respectively and consider each term separately,

\paragraph{$S_0$:}%
With our $\A^\f$, $\text{CS}(\A^\f)$ is quadratic because $\A^\f_{\bar\spa}=0$.  So, using $\p_{\bar\spa}\A^\f  =\p_{\bar\spa}\A'$, we see
\begin{eqaligned}
    \mathrm{CS}(\A^\f) &= \tr \A^\f   \wedge \rd\bar \spa \p_{\bar \spa } \A^\f = \tr \A^\f \wedge \rd\bar \spa \p_{\bar \spa } \A' = \tr\big(\A'  \wedge \rd\bar \spa \p_{\bar\spa} \A'\big) + \rd\tr\big((A +\spa^{-1}\varphi + \spa\wt\varphi)\wedge \A'\big)\,.
\end{eqaligned}
However, the boundary conditions on $\A' $ imply that the last term vanishes at $\spa=0,\infty$ so that even when multiplied by $\omega$ it extends as an exact form to $\CP^1$. As such, it integrates to zero so that 
\begin{equation}
    S_0 = \frac{\im}{8\pi^2}\bigintsss\omega\wedge \tr(\A'  \wedge \rd\bar \spa \p_{\bar \spa } \A' )\,.
\end{equation}
To simplify the discussion for $S_1$ and $S_2$, we use the $\gamma$ gauge freedom in \eqref{eq:gauge freedom in f} to set $f_0=\text{id}$ so that $f_\infty=h^{-1}$ (see \eqref{eq:definition of f_0 and f_infty} and \eqref{h-def}) and 
\begin{eqaligned}
(A,\varphi,\wt \varphi)=(h^{-1}\p h, h^{-1}(\p \psi ) h,\dbar\wt \psi)\,.
\end{eqaligned}

\paragraph{$S_1$:} %
We first note that the boundary conditions on $\A' $ imply that the contribution from $\omega\wedge \rd(J A_0)$ is smooth at $\spa=0,\infty$, so is exact and integrates to zero over $\CP^1$. We are left with the boundary integral 
\begin{equation}
    S_1 = \frac{\im}{8\pi^2}\lim_{\epsilon\to 0}\bigintsss_{\Sigma\times\p U_\epsilon} \omega \wedge\tr \left(J\wedge(A + \spa^{-1}\varphi +\spa\wt \varphi)\right)\,.
\end{equation}
This is essentially a residue calculation on $\CP^1$ with second-order poles at $\spa=0,\infty$ multiplying $\varphi,\wt\varphi$.  This picks out the $\spa$-derivative of $J$ at these points, together with simple poles multiplying $A$.  Noting the opposite signs coming from $\spa=0,\infty$ and using \eqref{J-def} and \eqref{psi-def}, we obtain
\begin{eqaligned}
    &S_1 =  \frac{1}{4\pi}\bigintsss_\Sigma\tr\big(\p_zJ|_{z=0}\wedge\varphi + z^2\p_z J|_{z=\infty}\wedge\wt\varphi - J|_{z=\infty}\wedge A\big) \\
    &= \frac{1}{4\pi}\bigintsss_\Sigma \tr\big(- (\rd\tilde\psi)\wedge\varphi + h^{-1}\rd\psi h\wedge\tilde\varphi + h^{-1}\rd h\wedge A\big) \\
    &= \frac{1}{4\pi}\bigintsss_\Sigma \tr\big(h^{-1}\bar\p h\wedge h^{-1}\p h - 2\dbar \wt \psi \wedge h^{-1} (\p\psi ) h\big)\,.
\end{eqaligned}
Comparison with \eqref{eq:action for Hitchin equations} shows that it can be recognized as $\tfrac{\im}{4\pi}S_\text{H}[h,\psi,\tilde\psi]$ without the WZ term.

\paragraph{$S_2$:}
Note first that $\omega=\rd\log\spa$ where we take the branch of $\log \spa$ so that the cut is along the real axis with jump $2\pi \im$.  Secondly, since  $\rd\tr (J^3)=0$, the integrand is exact $\rd (\log \spa \,\tr J^3)$.  Thus, by Stokes theorem on $\CP^1-\R_+$ with the cut along the positive real axis, we arrive at 
\begin{equation}
    S_2 = - \frac{\im}{24\pi^2} \lim_{\epsilon\to 0} \bigintsss_{\Sigma\times U_\epsilon} \rd (\log \spa \, \tr J^3)= - \frac{1}{12\pi}\bigintsss_{\Sigma\times\R^+} \tr J^3\,.
\end{equation}
Since $f^{-1}$ smoothly interpolates from the identity at $\spa=0$ to $h$ at $\spa=\infty$, we recognize this as the WZ term $\tfrac{\im}{4\pi}I_\text{WZ}[h]$ in \eqref{eq:action for Hitchin equations}.\footnote{Rewriting the Wess--Zumino term using $f\rd f^{-1}$ generates a further sign, which is cancelled by reversing the orientation of $\R^+$.}

\smallskip Thus, we see that the modes of $\A'$ decouple and have no zero modes, and so we can integrate them out, leaving us with $S_1+S_2$ yielding our action for the Hitchin equations \eqref{eq:action for Hitchin equations} up to an overall constant.\qed 
\end{prf}

\paragraph{Gauge Invariance.}  In our proof above, gauge invariance is only broken when we make the choice $f_0=I$.  We can relax this condition, allowing a non-trivial $f_0$, and when reality conditions are imposed, a natural restriction is one in which  $h$ is taken to be in a constant standard form for the Hermitian metric, i.e., the identity matrix when positive definite.  The gauge connection $A$ is then skew-Hermitian.  With this choice, our variables become
\begin{equation}
    (A,\Psi,\wt\Psi)= (-\dbar f_0 f_0^{-1}- \p f_\infty f_\infty^{-1}, f_\infty \psi f_\infty^{-1},f_0 \wt \psi f_0^{-1})\,,
\end{equation}
where the Higgs field is
\begin{equation}
(\varphi,\wt\varphi) = (\p_A\Psi,\dbar_A\wt\Psi)\,.
\end{equation}
The gauge freedom $\delta(\Psi,\wt\Psi)=(\wt\chi,\chi)$ now satisfies $\p_A\chi=0=\dbar_A\wt\chi$ so that $(\varphi,\wt\varphi)$ are invariant.

\smallskip The action \eqref{eq:action for Hitchin's equations, introduction} can be expressed in terms of the gauge-covariant data $f_0,f_\infty,\Psi,\wt\Psi$ by substituting $h = f_\infty^{-1}f_0$, $\psi = f_\infty^{-1}\Psi f_\infty$ and $\wt\psi = f_0^{-1}\wt\Psi f_0$, yielding
\be \begin{aligned} \label{eq:unitary-action}
&S_\mathrm{H}[f_0,f_\infty,\Psi,\wt\Psi] = \im\int_\Sigma\tr(j_0^{1,0}\wedge j_0^{0,1} - 2j_\infty^{1,0}\wedge j_0^{0,1} + j_\infty^{1,0}\wedge j_\infty^{0,1}) + I_\mathrm{WZ}[\f_\infty,\f_0] + \int_\Sigma|\p_A\Psi|^2\,.
\end{aligned} \ee
Here $j_0 = \d f_0 f_0^{-1}, j_\infty = \d f_\infty f_\infty^{-1}$.  The norm $|\cdot|$ is now defined using the standard constant Hermitian metric, and the Wess--Zumino term takes the same form as in equation \eqref{eq:WZ-term} though now defined using a homotopy $f:\Sigma\times[0,1]$ from $f_0$ on $\Sigma\times0$ to $f_\infty$ on $\Sigma\times1$.

\smallskip In a general unitary gauge, $f_0$ must be nontrivial with $f_\infty^*f_0 = I$.  These reality conditions are compatible with unitary $\gamma$-gauge transformations, under which the action \eqref{eq:unitary-action} is invariant.

\subsection{Symplectic Structure, and Hitchin Hamiltonians}\label{sec:symplectic structure, and Hitchin Hamiltonians}

We have derived an action \eqref{eq:action for Hitchin equations} whose equations of motion are Hitchin's equations. On the other hand, $\hit$, the space of solutions of Hitchin's equations, is a hyperk\"ahler manifold which has three distinguished complex structures $\mcal{I},\mcal{J},$ and $\mcal{K}$. Therefore, one may wonder how to derive these symplectic structures from the perspective of 4d Chern--Simons theory. We first show that the symplectic structure derived from 4d Chern--Simons theory with $\omega$ as in \eqref{eq:omega for the complex structure I} is the canonical symplectic structure associated with the complex structure $\mcal{I}$, and is given by 
\be \label{eq:symplectic structure in the complex structure I}
\Omega_\mathcal{I} = -\frac{1}{4\pi}\bigintsss_\Sigma\tr\left(\delta A\wedge \delta A + \delta\phi\wedge\delta\phi\right) = -\frac{\im}{2\pi}\bigintsss_{\Sigma}|\rd^2w|\,\tr\left(\delta A_{\bw}\wedge\delta A_w + \delta\phi_{\bw}\wedge\delta\phi_w\right)\,,
\ee
where $|\rd^2w|$ denotes the measure corresponding to the type-$(1,1)$ form $\im\,\rd w\wedge\rd\bw$.\footnote{Our convention for the canonical symplectic structures on $\hit$, given in \eqref{eq:symplectic structure in the complex structure I} and \eqref{eq:symplectic structures in the complex structures J and K}, are the same as \cite[eq. (4.8)]{KapustinWitten200604}. The difference is that $\phi$ is Hermitian here while it is skew-Hermitian in \cite{KapustinWitten200604}. Therefore, there are some sign differences. For example, the relative sign of \eqref{eq:symplectic structure in the complex structure I} is plus ($+$) while it is minus ($-$) in \cite[eq. (4.8)]{KapustinWitten200604}. These signs ensure the reality of the corresponding symplectic structures.} Furthermore, we explain how to construct Hitchin Hamiltonians in terms of four-dimensional gauge fields. 

\paragraph{Symplectic Structure.} To compute the symplectic structure from 4d Chern--Simons theory, we first decompose $\CP^1-\{0,\infty\}\simeq \R^+\times S^1$, where $\R^+$ factor can be interpreted as the `time' direction. This yields the symplectic structure\footnote{The pre-symplectic structure on the space of solutions of an action can be obtained using the covariant phase space formalism \cite{KijowskiSzczyrba197606,CrnkovicWitten198704,LeeWald199003,Takens1977,Zuckerman198709}. First, one constructs the so-called pre-symplectic potential $\Theta$ from the variation of the action as $\delta S=\text{E.O.M}+\Theta$. Then, the symplectic structure is given by $\Omega=\delta\Theta$. A pre-symplectic potential which is non-degenerate will then define a symplectic potential.} on the space of solutions of field equations as
\begin{equation}\label{eq:symplectic structure in 4d CS for omega(0,0)}
    \Omega_{\text{CS}_4} = \frac{\im}{8\pi^2}\bigintsss_\Sigma\bigointsss_{S^1}\frac{\rd z}{z}\wedge\tr(\delta\A\wedge\delta\A)\,.
\end{equation}
As we explained around \eqref{eq:connection A0}, by a judicious gauge choice, the gauge connection $\A$ is gauge-equivalent to the Lax form of the Hitchin system, given in \eqref{eq:Lax-Hitchin}. We have
\begin{equation} \label{eq:expression of delta A in 4d CS}
    \delta\A = f^{-1}(\delta\A^f) f - f^{-1}D_\A(\delta ff^{-1})f\,.
\end{equation}
Since $\delta\A$ is the on-shell variation it obeys $\tfrac{\rd z}{z}\wedge D_{\A^f}(\delta\A)^f = 0$.  As a result
\be \Omega_{\text{CS}_4} = \frac{\im}{8\pi^2}\int_\Sigma\oint_{S^1}\frac{\rd z}{z}\wedge\Big(\tr(\delta\A^f\wedge\delta\A^f) - 2\rd\tr\big(\delta\A^f\wedge\delta f f^{-1}\big) + \rd\tr\big(\delta ff^{-1}\wedge D_{\A^f}(\delta ff^{-1})\big)\Big)\,. \ee
Neglecting contributions from boundaries in $\Sigma$, we therefore find that
\be \label{eq:symplectic structure after gauge transformation} \Omega_{\text{CS}_4} = \frac{\im}{8\pi^2}\int_\Sigma\oint_{S^1}\frac{\rd z}{z}\wedge\tr(\delta\A^f\wedge\delta\A^f)\,. \ee
Putting the expression for the Hitchin Lax \eqref{eq:Lax-Hitchin} into \eqref{eq:symplectic structure after gauge transformation} gives
\begin{eqaligned}
     \Omega_{\text{CS}_4} = - \frac{1}{4\pi}\bigintsss_\Sigma\tr\left(\delta A\wedge\delta A + 2\delta\varphi\wedge\delta\varphi^*\right) = - \frac{1}{4\pi}\bigintsss_\Sigma\tr\left(\delta A\wedge\delta A + \delta\phi\wedge\delta\phi\right)\,,
\end{eqaligned}
where we have used $\phi=\varphi+\varphi^*$ in the second line. We immediately recognize that $\Omega_{\text{CS}_4}$ coincides with $\Omega_{\mcal{I}}$, given in \eqref{eq:symplectic structure in the complex structure I}. Thus the space of solutions of the field equations of the action \eqref{eq:action for Hitchin equations} leads to the symplectic structure that is the K\"ahler form for $\hit$ as a complex manifold in the complex structure $\mcal{I}$.  This is the complex structure that arises from the identification of $\hit$ with the moduli space of Higgs bundles.  We now elaborate further on this.

\paragraph{$\hit$ as an Algebraically Completely Integrable System and Hitchin  Hamiltonians.}
In  \cite{Hitchin198701}, Hitchin shows that for compact $\Sigma$ of genus $g\geq 2$, the space of solutions of the action \eqref{eq:action for Hitchin equations} is essentially the moduli space $\mscr{M}_{\text{Hig}}^{\text{s}}(n,d)$ of Higgs bundles, i.e., pairs $(E,\varphi)$ where $E$ is a holomorphic vector bundle and $\varphi$ the holomorphic $(1,0)$-form with values in $\End(E)$ introduced above.  He goes on to show that this moduli space can be realized as an algebraically completely integrable Hamiltonian system.  We sketch the construction very briefly as follows. The details can be found in \cite{Hitchin198701}.

\smallskip Firstly general hyperk\"ahler lore shows that the complex 2-form  $\Omega_{\mathcal{I}}+\im \Omega_{\mathcal{J}}$ is holomorphic symplectic  with respect to the complex structure $\mathcal{I}$. This can be represented explicitly as 
\begin{equation}
    \Omega_{\mcal{J}}+\im\,\Omega_{\mcal{K}}
=-\frac{1}{2\pi\im}\bigintsss_\Sigma(\delta\phi\wedge(1-\im\star)\delta A)=\frac{1}{\pi}\bigintsss_\Sigma\,|\rd^2w|\tr(\delta\phi_w\wedge\delta A_{\bw}).
\end{equation}
where $|\rd^2w|$ denotes the measure corresponding to the type-$(1,1)$ form $\im\,\rd w\wedge\rd\bw$.
This expression  follows from the representations 
    \begin{eqaligned}\label{eq:symplectic structures in the complex structures J and K}
    \Omega_{\mathcal{J}}&=\frac{1}{2\pi}\bigintsss_\Sigma\tr(\delta\phi\wedge\star\delta A)=\frac{1}{2\pi}\bigintsss_{\Sigma}|\rd^2w|\tr\left(\delta\phi_{\bw}\wedge\delta A_w+\delta\phi_w\wedge\delta A_{\bw}\right),
    \\
    \Omega_{\mathcal{K}}&=\frac{1}{2\pi}\bigintsss_\Sigma\tr(\delta\phi\wedge\delta A)=\frac{\im}{2\pi}\bigintsss_{\Sigma}|\rd^2w|\tr\left(\delta\phi_{\bw}\wedge\delta A_w-\delta\phi_w\wedge\delta A_{\bw}\right),
\end{eqaligned}
for the canonical symplectic structures associated with the complex structures $\mcal{J}$ and $\mcal{K}$.    

\smallskip We must then present the complete family of holomorphic commuting Hamiltonians to realize this as an algebraically completely integrable system.
These can be presented via the Hitchin fibration
    \begin{equation}
        \mbf{h}:\mscr{M}_{\text{Hig}}^{\text{s}}(n,d)\to\mbf{B},
    \end{equation}
    which sends a Higgs bundle $(E,\varphi)$ to the coefficients of the characteristic polynomial $\det(\lambda\mds{1}-\varphi)$. For the group $G_\C=\text{SL}(n,\C)$, these coefficients are the invariant polynomials 
    \begin{equation}\label{eq:G-invariant polynomials for SL(n,C)}
        P_m(\varphi)\coloneqq\tr(\varphi^m)\in H^0(\Sigma,K^{\otimes m}), \qquad m=2,\ldots,n,
    \end{equation}
    where $K$ denotes the canonical bundle of $\Sigma$. As such, the vector space $\mbf{B}$, the Hitchin base, is  
    \begin{equation}
        \mbf{B}=\bigoplus_{m=2}^nH^0(\Sigma,K^{\otimes m}).
    \end{equation}
An application of Riemann--Roch Theorem gives $\dim_\C\mbf{B}=(g-1)(n^2-1)$. The crucial point is that there exists a dense open subset $\mbf{B}'\subset\mbf{B}$ such that $\mbf{h}^{-1}(\mbf{B}')\subset\mscr{M}_{\text{Hig}}^{\text{s}}(n,d)$  is a complex Lagrangian torus with respect to the complex-symplectic structure $\Omega_{\mcal{J}}+\im\,\Omega_{\mcal{K}},
$.

\smallskip To construct the corresponding Hamiltonians, we pick $v\in H^{(0,1)}(\Sigma,K^{\otimes(1-m)})\simeq H^1(\Sigma,K^{\otimes(1-m)})$, and set
\begin{equation}\label{eq:Hitchin Hamiltonians}
    H_{v,m}\coloneqq\bigintsss_\Sigma v P_m(\varphi), \qquad m=2,\ldots,n.
\end{equation}
The integrand is now a $(1,1)$-form, so it can be integrated over $\Sigma$. The number of independent Hamiltonians corresponds to the number of independent $v$s, which can be computed by the Riemann--Roch Theorem.  This gives for $g\geq 2$
\begin{equation}
    \sum_{m=2}^{n-1}\dim_{\C}H^1(\Sigma,K^{\otimes(1-m)})=(n^2-1)(g-1)=\dim_\C\mbf{B}.
\end{equation}
This counting can be used to show that we have a complete set of Hamiltonians for a completely integrable system.
The  $H_{v,m}$s in \eqref{eq:Hitchin Hamiltonians} do not depend on $A$, and so obviously satisfy
\begin{equation}
    \{H_{v,m},H_{v',m'}\}=0,\qquad \forall m,m'=2,\ldots,n,
\end{equation}
where $\{\cdot,\cdot\}$ is defined with respect to the holomorphic-symplectic form $\Omega_{\mcal{J}}+\im\Omega_{\mcal{K}}$, and $v,v'$ belong to $H^1(\Sigma,K^{\otimes(1-m)})$ and $H^1(\Sigma,K^{\otimes(1-m')})$, respectively. Hence, we have $\dim_\C\mbf{B}$ Hamiltonians in involution. As $\dim_{\C}\mbf{B}=\frac{1}{2}\dim_{\C}\mbf{h}^{-1}(\mbf{B}')$, by the principle of Liouville integrability, the very existence of the Hamiltonians \eqref{eq:Hitchin Hamiltonians} which are in involution makes the integrability of $\mbf{h}^{-1}(\mbf{B}')$ manifest.

\smallskip Therefore, to construct Hitchin Hamiltonians from 4d Chern--Simons theory, we just need to know how to construct the Higgs field from the four-dimensional gauge fields. By \eqref{eq:gauge field cA in the holomorphic gauge} and the discussion around it, this is a trivial matter: we first go to the holomorphic gauge \eqref{Hol-gauge} in which the four-dimensional gauge field can be written as \eqref{eq:gauge field cA in the holomorphic gauge}. Then, the Higgs field is given by
\begin{equation}
    \varphi=\text{Res}_{\spa=0}(\mcal{A})=\frac{1}{2\pi\im}\bigointsss_{|\spa|=\epsilon}\mcal{A},
\end{equation}
Combining this with \eqref{eq:Hitchin Hamiltonians}, we have the following simple explicit formula for the Hitchin Hamiltonians in terms of the gauge fields of four-dimensional Chern--Simons theory
\begin{equation}
    H_{v,m}=\bigintsss_\Sigma vP_m(\text{Res}_{\spa =0}(\mcal{A})).
\end{equation}
Thus, we have constructed the Hitchin Hamiltonians from four-dimensional Chern--Simons theory.

\section{\texorpdfstring{$\CP^1$}{CP1}-Family of 2d Actions Associated with Hitchin's equations}
\label{sec:CP^1-family of 2d actions associated with Hitchin's equations}
In \S\ref{sec:from 4d CS theory to Hitchin's equations}, we described an action whose field equations are Hitchin's equations, and connected the space of solutions of its field equations with the moduli space of Higgs bundles. On the other hand, $\hit$, as a hyperk\"ahler manifold, has a whole $\CP^1$-family of symplectic structures \eqref{eq:family of symplectic structure on Hitchin moduli space}. Therefore, a natural follow-up question is whether there exists a corresponding family of 2d actions such that the space of solutions of their field equations is related to $\hit$ as a complex manifold in the complex structure $\mathcal{J}_{\zeta}$, given in \eqref{eq:family of symplectic structure on Hitchin moduli space}.

\smallskip The purpose of the present section is to answer this question in the affirmative. More specifically, after explaining the relevant 4d Chern--Simons theory setup, we construct in \S\ref{sec:CP^1 family of 4d CS theories and the corresponding 2d actions} a $\CP^1$-family of two-dimensional actions parametrized by $\zeta \in\CP^1$. We then show in \S\ref{sec:hyperkahler family of symplectic structures} that the symplectic structures on the spaces of solutions to their equations of motion coincide with the symplectic structure in the complex structure $\mcal{J}_\zeta$.

\subsection{\texorpdfstring{$\CP^1$}{}-Family of 4d CS Theories and the Corresponding 2d Actions}\label{sec:CP^1 family of 4d CS theories and the corresponding 2d actions}

Since there is a whole $\CP^1$ family of hyperk\"ahler structures on $\hit$, it is expected from our computations in \S\ref{sec:symplectic structure, and Hitchin Hamiltonians} that there exists a family of 4d Chern--Simons theories that produces this hyperk\"ahler family of symplectic structures. For this matter, we propose the following $\CP^1$-family of meromorphic $(1,0)$-form on $C$
\begin{equation}\label{eq:CP^1 family of omegas}
    \omega(\zeta ,\bar{\zeta})\coloneqq \frac{(\zeta +1/ \bar\zeta )^2\spa \rd \spa}{(\spa-\zeta )^2 (\spa +1/\bar\zeta )^2}, \qquad \zeta \in\CP^1. 
\end{equation}
First note that
\begin{equation}
    \lim_{\zeta \to0}\omega(\zeta ,\bar{\zeta})=\frac{\rd \spa}{\spa},
\end{equation}
giving the meromorphic $(1,0)$-form used in \S\ref{sec:4dCS-Hitchin} that led to the 4d Chern--Simons theory formulation of Hitchin's equations with symplectic structure $\Omega_{\mcal{I}}$. We will see that the $\zeta $ above will be the twistor parameter $\zeta$ of the Hitchin hyperk\"ahler structure, parametrizing the symplectic structures in \eqref{eq:family of symplectic structure on Hitchin moduli space}.  Here we will work with complexified fields; Euclidean reality conditions can be imposed in direct analogy with section \ref{sec:4dCS-Hitchin}.

\smallskip We will derive the corresponding family of 2d actions parametrized by $\CP^1$ following the procedure of the previous section. We must first introduce appropriate boundary conditions. Our $\omega(\zeta,\bar{\zeta})$ of \eqref{eq:CP^1 family of omegas} has simple zeroes at $z=0,\infty$, and double poles at $z=\zeta ,-1/\bar{\zeta}$. At the zeroes of $\omega(\zeta ,\bar{\zeta})$, we allow the following poles
\begin{equation}
    \begin{alignedat}{3}
    \A_w&\sim\mathcal{O}(\spa^{-1})\,, &\qquad \spa&\to 0\,,
    \\
    \A_{\bw}&\sim\mathcal{O}(\spa)\,, &\qquad \spa&\to\infty\,.
\end{alignedat}
\end{equation}
 At the poles of $\omega(\zeta,\bar{\zeta})$ we impose the boundary conditions
\begin{equation} \label{eq:zeta-A-bcs}
    \begin{alignedat}{3}
        \A_{\bw}&\sim\mathcal{O}((\spa-\zeta)^2)\,, &\qquad z&\to\zeta\,,
        \\
        \A_{w}&\sim\mathcal{O}((\spa+\bar{\zeta}^{-1})^2)\,, &\qquad z&\to - \bar\zeta^{-1}\,.
    \end{alignedat}
\end{equation}
These boundary conditions coalesce to those in \eqref{A-bc} as $\zeta\to 0$.  We will find it useful to write $\widehat\zeta = - 1/\bar\zeta$ for the stereographic coordinate of the point antipodal to $\zeta$.

\smallskip The derivation of the 2d action follows the procedure in \S\ref{sec:4dCS-Hitchin}. We still need to go to a gauge in which \eqref{Hol-gauge} is satisfied, but instead of \eqref{eq:definition of f_0 and f_infty} we define the frame fields
\begin{equation}
    f_\zeta\coloneqq f\big|_{\spa=\zeta}\,,\qquad f_{\widehat\zeta}\coloneqq f\big|_{\spa=\widehat\zeta}\,,
\end{equation}
admitting actions on the left by $\gamma$-gauge transformations.  We can again form invariant `holomorphic Wilson lines' $k = f_{\widehat\zeta}^{-1}f_\zeta$.

\smallskip We will find that the gauge connection $A$ and Higgs field $\phi$ are fixed in terms of potentials $(k,\eta,\wt\eta)$ determined by the asymptotics of $f$, but now in neighbourhoods of $z=\zeta,\widehat\zeta$.\footnote{We use different notation for the holomorphic Wilson line $k$ and potentials $\eta,\widetilde\eta$ than in \S\ref{sec:from 4d CS theory to Hitchin's equations}.  This is emphasise that $k$ is not a frame for the connection $A$, but rather for $A_\zeta = A + \zeta\wt\varphi - \bar\zeta\varphi$, and furthermore that the potentials $\eta,\wt\eta$ are defined in a holomorphic gauge for this connection.} Indeed,
\begin{eqaligned} \label{eq:zeta-psi-def}
f&= f_\zeta\big(1-(z-\zeta)\wt\eta + \mcal{O}\big((z-\zeta)^2\big)\big)\,, &\qquad\text{as}\qquad z&\to\zeta\,, \\ 
f&=f_{\widehat\zeta}\big(1 - (z^{-1} - \widehat\zeta^{-1})\eta + \mcal{O}\big((z-\widehat\zeta)^2\big)\big)\,, &\qquad \text{as}\qquad  z&\rightarrow\widehat\zeta\,.
\end{eqaligned}
The fields $(f_\zeta,f_{\widehat\zeta},\eta,\tilde\eta)$ transform under residual gauge transformations, which straightforwardly generalize equations \eqref{psi-gauge}.  To determine $(A,\phi)$ we fix the gauge under $\gamma$-transformations by setting $f_\zeta=1$, so that $k = f_{\widehat\zeta}^{-1}$, and write
\be J^\prime \coloneqq J + z\wt\varphi + A + \frac{1}{z}\varphi \ee
where $J = \rd ff^{-1}$ and $J^\prime$ obeys the boundary conditions \eqref{eq:zeta-A-bcs}.  Then the gauge field $A$ appearing in Hitchin's equations is fixed to
\be \label{eq:zeta-A} A = k^{-1}\p k - \widehat\zeta^{-1}k^{-1}(\p\eta)k - \zeta\dbar\wt\eta \ee
while the $1$-form $\phi$ is
\be \label{eq:zeta-phi} \phi = k^{-1}(\p\eta)k + \dbar\wt\eta\,. \ee

\smallskip To determine the resulting 2d action, we use the same procedure as explained in the proof of Lemma \ref{lem:2d action from 4d cs action}, which yields $\frac{\im}{4\pi}S_{\mathrm{H},\zeta}$ where
\begin{equation} \label{eq:action for the complex structure J_{zeta_0}}
    \begin{alignedat}{2}
    &S_{\text H,\zeta}[k,\eta,\widetilde\eta] = \frac{1-|\zeta|^2}{1+|\zeta|^2}S_\text{WZW}[k] - 2\im(1+|\zeta |^2)\bigintsss_\Sigma\tr\big(k^{-1}\p\eta k\wedge \dbar\wt\eta\big) \\
    &-2\im\zeta\bigintsss_\Sigma\tr(k^{-1}\p k\wedge\,\dbar\wt\eta) - 2\im\bar\zeta\int_\Sigma\tr(\p\eta\wedge\dbar kk^{-1})\,,
    \end{alignedat}
\end{equation}
where $S_\text{WZW}[k]$ is the WZW action given in equation \eqref{eq:action for Hitchin equations}.  It is straightforward to construct a version of this action depending on gauge covariant data by substituting
\be (k,\eta,\wt\eta) = (f_{\wh\zeta}^{-1}f_\zeta,f_{\hat\zeta}^{-1}\boldsymbol{\eta}f_{\hat\zeta},f_\zeta^{-1}\wt{\boldsymbol{\eta}}f_\zeta)\,. \ee
The result is a $\zeta$-dependent counterpart of \eqref{eq:unitary-action}.

\smallskip Writing $A_\zeta = A + \zeta\wt\varphi + \widehat\zeta^{-1}\varphi = k^{-1}\p k$ for the Chern connection associated to $k$, and $\cD_\zeta = \p + A_\zeta$ for the associated partial connection, the equation of motion obtained by varying $k$ can be written as
\be \frac{1-|\zeta|^2}{1+|\zeta|^2}F(A_\zeta) + (1+|\zeta|^2)[\varphi,\widetilde\varphi] - \zeta\dbar\varphi + \bar\zeta\cD_\zeta\wt\varphi = 0\,. \ee
Varying $\eta$ and $\wt\eta$ we find
\be (1+|\zeta|^2)\cD_\zeta\wt\varphi - \zeta F(A_\zeta) = 0\,,\qquad (1+|\zeta|^2)\dbar\varphi + \bar\zeta F(A_\zeta) = 0\,. \ee
By taking linear combinations of the above, it's straightforward to see that
\be
F(A_\zeta - \zeta\wt\varphi + \bar\zeta\varphi) + [\varphi,\wt\varphi] = 0\,, \qquad \cD_\zeta\wt\varphi + \bar\zeta[\varphi,\wt\varphi] = 0\,,\qquad \bar\dbar\varphi - \zeta[\varphi,\wt\varphi] = 0\,.
\ee
We recognise these as Hitchin's equations expressed in terms of $A_\zeta,\varphi,\wt\varphi$ and in a holomorphic gauge for $A_\zeta$.

\paragraph{Relation to Harmonic Maps.} 
The 4d Chern--Simons reduction to Harmonic maps or principal chiral models are treated in detail in \cite{CostelloYamazaki201908}. We first remark that this also uses an $\omega$ of the form \eqref{eq:CP^1 family of omegas}, but the boundary conditions used are to instead impose $\A=0$ at $\zeta$ and $-1/\bar\zeta$ instead of  \eqref{eq:zeta-A-bcs}.  The Wilson line from $\zeta$ to $-1/\bar\zeta$  then gives the (twisted) harmonic maps of \cite{Donaldson198707,Corlette198801}.  The Lax pair thereby obtained are gauge equivalent to ours, but the twisted harmonic map is realized in a different gauge from our $k$.  \qed 

\subsection{Hyperk\"ahler Family of Symplectic Structures}\label{sec:hyperkahler family of symplectic structures}

In \S\ref{sec:CP^1 family of 4d CS theories and the corresponding 2d actions}, we derive a family of 2d actions \eqref{eq:action for the complex structure J_{zeta_0}} parametrized by $\zeta \in\CP^1$ whose equations of motion, for each value of $\zeta $, are Hitchin's equations \eqref{eq:Hitchin's equations, introduction}. However, our aim was to find the relationship between $\zeta $ and the twistor parameter of the Hitchin moduli space $\zeta$, which determines the family of symplectic structures \eqref{eq:family of symplectic structure on Hitchin moduli space}. To this aim, we proceed as we did in \S\ref{sec:symplectic structure, and Hitchin Hamiltonians} and compute the symplectic structure that 4d Chern--Simons theory with the choice \eqref{eq:CP^1 family of omegas} associates to $\Sigma\times S^1$, as discussed in \S\ref{sec:symplectic structure, and Hitchin Hamiltonians}. 

\smallskip To compute the symplectic structure, we simply replace \eqref{eq:symplectic structure in 4d CS for omega(0,0)} with 
\begin{equation} \label{eq:4dCS-zeta-symplectic}
    \Omega_{\text{CS}_4}[\zeta,\bar{\zeta}] = \frac{\im}{8\pi^2}\bigintsss_\Sigma\bigointsss_{S^1}\omega(\zeta ,\bar{\zeta})\wedge\tr(\delta\A\wedge\delta\A).
\end{equation}
where $\omega(\zeta ,\bar{\zeta})$ is given in \eqref{eq:CP^1 family of omegas}. By what we have explained in \S\ref{sec:CP^1 family of 4d CS theories and the corresponding 2d actions}, and by a judicious gauge choice, we can make $\A$ to coincide with the Lax form of the Hitchin system, as given in \eqref{eq:Lax-Hitchin}. As such, $\delta\A$ is given by \eqref{eq:expression of delta A in 4d CS}. We still need to decide on the choice of the contour $S^1$. It is easy to check that if we take $S^1$ to be a small circle around $z=0$ or $z = \infty$, then the integral vanishes. This is because $\omega(\zeta , \bar{\zeta})$ provides a compensating factor for the poles in $\delta\A \wedge \delta\A$. Therefore, $ S^1$ should be chosen as a contour that separates $z = \zeta $ from $z = -1/\bar{\zeta}$. For the purpose of the computation, we may use either $z = \zeta $ or $z = -1/\bar{\zeta}$. Since there is no contribution from $z = 0$ or  $z = \infty$, the contour can be deformed across these points without issue, and the results obtained by taking $z = \zeta $ or $z = -1/\bar{\zeta}$ agree up to an overall sign arising from the orientation of the contour. Therefore, without loss of generality, we choose $S^1$ to be a small circle around $z=\zeta $.

\smallskip Substituting \eqref{eq:expression of delta A in 4d CS} into \eqref{eq:4dCS-zeta-symplectic}, repeating the argument of section \ref{sec:symplectic structure, and Hitchin Hamiltonians} and computing the resulting integral gives
\begin{equation}
    \Omega_{\text{CS}_4}(\zeta,\bar{\zeta})= - \frac{1}{2\pi(1+|\zeta |^2)}\bigintsss_\Sigma\tr\left(\frac{1}{2}(1-|\zeta |^2)(\delta A\wedge\delta A-2\delta\varphi\wedge\delta\varphi^*)-2\bar{\zeta}\delta \varphi\wedge\delta A-2\zeta \delta\varphi^*\wedge\delta A\right)\,.
\end{equation}
Then, \eqref{eq:symplectic structure in the complex structure I}, we can write the first term as
\begin{equation}
    \frac{1}{2}\bigintsss_\Sigma\tr(\delta A\wedge\delta A-\delta\varphi\wedge\delta\varphi^*)=-2\pi\,\Omega_{\mcal{I}}\,,
\end{equation}
and from \eqref{eq:symplectic structures in the complex structures J and K}, we have
\begin{equation}
    -2\bigintsss_\Sigma\tr\left(\delta\varphi\wedge\delta A\right) = +2\pi\im(\Omega_{\mcal{J}}+\im\Omega_{\mcal{K}})\,,\qquad -2\bigintsss_\Sigma\tr\left(\delta\varphi^*\wedge\delta A\right) = -2\pi\im(\Omega_{\mcal{J}}-\im\Omega_{\mcal{K}})\,.
\end{equation}
Hence, we arrive at
\begin{equation}\label{eq:CP^1 family of symplectic structures from 4d CS, first form}
    \Omega_{\text{CS}_4}(\zeta ,\bar{\zeta})=\frac{1-|\zeta |^2}{1+|\zeta |^2}\Omega_{\mcal{I}}+\im\frac{\zeta -\bar{\zeta}}{1+|\zeta |^2}\Omega_{\mcal{J}}+\frac{\zeta +\bar{\zeta}}{1+|\zeta |^2}\Omega_{\mcal{K}}.
\end{equation}
The comparison between the coefficients of the canonical symplectic structure $\Omega_{\mcal{I}},\Omega_{\mcal{J}},$ and $\Omega_{\mcal{K}}$ in \eqref{eq:CP^1 family of symplectic structures from 4d CS, first form} and \eqref{eq:family of symplectic structure on Hitchin moduli space} shows that starting with 4d Chern--Simons theory with the choice \eqref{eq:CP^1 family of omegas} of meromorphic $(1,0)$-form on $\CP^1$, we precisely land on the symplectic structure associated with the complex structure $\mcal{J}_{\zeta}$
\begin{equation}
    \Omega_{\text{CS}_4}(\zeta ,\bar{\zeta})=\Omega_{\mcal{J}_{\zeta }}.
\end{equation}
Therefore, we can identify the parameter $\zeta$ in \eqref{eq:CP^1 family of omegas} with the twistor parameter $\zeta$ of the Hitchin moduli space as appears in \eqref{eq:family of symplectic structure on Hitchin moduli space}.

\smallskip By varying $\zeta $ over $\CP^1$, \eqref{eq:CP^1 family of symplectic structures from 4d CS, first form} generates the whole $\CP^1$-family of  symplectic structures associated to the hyperk\"ahler structure on $\hit$. Our $\CP^1$-family of 4d Chern--Simons theory  using $\omega(\zeta,\bar\zeta)$ of \eqref{eq:CP^1 family of omegas} generates the $\CP^1$-family of 2d actions \eqref{eq:action for the complex structure J_{zeta_0}} giving the symplectic structures  $\Omega_{\mcal{J}_{\zeta }}$ on $\hit$, these being the K\"ahler 2-forms for the complex structures $\mcal{J}_{\zeta }$ on the space of solutions. 

\section{Affine Toda Theory}
\label{sec:affineToda}
The complexified Hitchin equations are, up to gauge, the generic reduction of the self-dual Yang--Mills from 4d to 2d along a non-degenerate pair of translations.  
Such reductions are described in \cite[\S 6.2]{MasonWoodhouse199605} and include a number of familiar 2d integrable systems, including harmonic maps, Toda field theory, and the sine-Gordon equation. These are distinguished as reductions by gauge choices and choices of `constants of integration' that naturally arise in the reductions.  When these models are reformulated as 4d Chern--Simons theories, these choices are encoded in the choice of $\omega$ and the boundary conditions on $\A$. It turns out that the harmonic map reductions require different such choices -- these are already well-understood in \cite{CostelloYamazaki201908} and require $\omega$ to have two double poles and two single zeroes with $\A$ vanishing at the double poles and allowed to have poles at the zeroes.

However, here we see that Toda field theories fit into the choices made for the Hitchin equations in \S\ref{sec:from 4d CS theory to Hitchin's equations} and can be understood as arising from the imposition of further symmetries.  This is a continuous symmetry for Toda field theory (here referring to Toda field theory defined using a Cartan matrix of finite type), but for affine Toda field theory (defined using a Cartan matrix of affine type), we instead impose a discrete symmetry.  Toda field theory is a CFT \cite{Braaten:1983pz,Gervais:1983ry,Mansfield:1982sq}, whereas affine Toda field theory is integrable but not conformal \cite{Mikhailov197910,LeznovSaveliev197909,MikhailovOlshanetskyPerelomov198007,OliveTurok198208}.

\begin{rmk}
    Let us briefly survey a number of connections between Chern--Simons theory and Toda field theories appearing in the literature.  Toda field theory can be realised as constrained, or equivalently as a gauged, WZW model \cite{Forgacs:1989ac,Balog:1990mu}.  This presentation has been exploited to engineer complex Toda theory in 3d Chern--Simons theory with Nahm pole boundary conditions \cite{Cordova:2016cmu}, and a lift to a `doubled' 4d Chern--Simons theory was identified in \cite{Stedman202009}. A 4d Chern--Simons realisation of affine Toda theory appears in \cite{FukushimaSakamotoYoshida202112} using the same meromorphic $1$-form $\omega$, but with order defects.  We anticipate that our treatment can be connected to theirs by integrating out local degrees of freedom living on the defects.  Homogeneous Sine-Gordon models, including in particular complex sine-Gordon, were engineered in \cite{ColeCullinanHoareLiniadoThompson202407} by reducing a gauged holomorphic Chern--Simons theory on twistor space to 4d Chern--Simons. Finally, we note that 4d Chern--Simons with Nahm pole boundary conditions in the topological direction was related to a 3d variant of Toda theory in \cite{AshwinkumarPngTan202008}. \qed 
\end{rmk}

\subsection{Toda Field Theories}

Toda theories can be defined for any Kac--Moody algebra, or equivalently for any generalised Cartan matrix, but for simplicity we restrict to the case of  $\mfk{su}(n)$ for the $A_{n-1}$ series and its untwisted affine counterpart $\widehat{\mfk{su}}(n)$.  Both cases can be treated together if we introduce variables $v_1,\ldots, v_n$ with $\sum_i v_i=0$.  We then have Lagrangians
\begin{equation} \label{eq:Toda-Lagrangian}
L = \sum_{i=1}^n 2\p_w v_i\p_{\bar w} v_i + \lambda \sum_{i=1}^{n-1} \e^{v_i - v_{i+1}}\,,\qquad \wh L= \sum_{i=1}^n \big(2\p_w v_i\p_{\bar w} v_i + \lambda \e^{v_i - v_{i+1}}\big)\,.
\end{equation}
In the affine case we take the indices to be cyclic $v_i = v_{n+i}$.  $L$ and $\wh L$ differ in the last interaction term $\lambda \e^{v_n-v_1}$ corresponding to the extra root in the affine Lie algebra.  The constraint $\sum_i v_i=0$ is consistent because the interactions depend only on the differences.  The equations of motion are
\be \label{eq:Toda-eom} \Delta v_i = \lambda\big(e^{v_i - v_{i+1}} - e^{v_{i-1} - v_i}\big)\,, \ee
where we understand $\exp v_0 =\exp v_{n+1} = 0$ in the finite case.\footnote{Both finite and affine Toda theories are often expressed in terms of alternative variables $t_i$ obeying $v_i = t_i - t_{i-1}$.  In the finite case $t_0 = t_n = 0$ whereas in the affine case $t_0 = t_n$ and $\sum_it_i = 0$.  They satisfy
$\Delta t_i = \lambda \exp\sum_jK_{ij}t_j$ for $K$ the Cartan matrix of $\mathfrak{su}(n)$ or $\wh{\mathfrak{su}}(n)$ as appropriate.}
The case $n=2$ contains the sinh/sine-Gordon equations.

\smallskip These systems are shown to be a reduction of Hitchin's equations in \cite[\S6.2]{MasonWoodhouse199605}.  This is done by imposing equivariance under a continuous symmetry in the finite case and a discrete symmetry in the affine case. In that work, they are understood as additional rotational symmetries in the reduction of the 4d self-dual Yang--Mills equations that act on the 2-plane of translational symmetry.  It naturally acts also on the spectral parameter, $\zeta\rightarrow \e^{\im\theta}\zeta$, and hence in our context, this symmetry naturally acts on our 4d-Chern--Simons theory too.  The symmetries are lifted to the bundle so as to act by the matrices  
\be \rho =  \begin{pmatrix}
1 & 0 & 0 & \dots & 0 \\
0 & \e^{\im\theta} & 0 & \dots & 0 \\
0 & 0 &  \e^{2\im\theta} & \ldots &0 \\
\vdots & \vdots & \vdots & \ddots & \vdots \\
0 & 0 & 0 & \dots & \e^{\im(n-1)\theta} \\
\end{pmatrix}\,,
\ee
at the fixed  points, with $\theta=2\pi\im/n$ in the discrete case.  This leads to the equivariance conditions:
\begin{equation}
\A(w,\e^{\im\theta}\zeta)=\rho\A(w,\zeta)\rho^{-1},
\end{equation}
which we take as the starting point for our reductions.

\subsection{Reduction to Two Dimensions}

We now show how these additional symmetries act on our reduction of 4d Chern--Simons to 2d actions so as to land on actions for affine Toda field theory.  Following through the reduction of 4dCS to the Hitchin system we introduce $f$ satisfying \eqref{eq:trivialise}
\begin{equation}
\p_{\bar \zeta} f=f\A_{\bar \zeta}\, , \qquad f_0=1\, . \label{f-def}
\end{equation}
 The equivariance of $\A$ and the uniqueness of solutions to \eqref{f-def} imply $f(w,\e^{\im\theta}\zeta)\rho=Af(w,\zeta)$ with $A$ independent of $\zeta$, but $\zeta=0$ is a  fixed point of the symmetry action with $f_0=1$ so $A=\rho$ and  we have
\begin{equation}
f (w,\e^{\im\theta}\zeta)=\rho f (w,\zeta)\rho^{-1}\, .
\end{equation}
Expanding this to first order at $\zeta=0,\infty$, using $h= f_\infty^{-1}$ we find
\begin{equation}
\rho h\rho^{-1}=h\, , \qquad \rho\psi\rho^{-1}=\e^{-\im\theta} \psi \, , \qquad  \rho\wt\psi\rho^{-1}=\e^{\im\theta}\wt \psi\, .
\end{equation}
Thus $h= \diag\{h_1,\ldots,h_n\} $ is diagonal, and since it also has unit determinant, we can write
\begin{equation}
    h_i= \e^{v_i}\,, \qquad \sum_i v_i=0\,.
\end{equation}
If we restrict $\theta$ to integer multiples of  $2\pi\im/n$ these conditions on $(\psi,\wt\psi)$ are equivalent to
\begin{equation}
\psi = \begin{pmatrix}
0 & \psi_1 & 0 & \dots & 0  \\
0 & 0 & \psi_2 & \dots & 0  \\
\vdots & \vdots & \vdots & \ddots & \vdots  \\
0 & 0 & 0 & \dots & \psi_{n-1} \\
\psi_n & 0 & 0 & \dots & 0  \\
\end{pmatrix} \, , \qquad \wt\psi = \begin{pmatrix}
0 & 0 &  \dots & 0 & \wt\psi_n \\
\wt\psi_1 & 0 &  \dots & 0 & 0 \\
0 & \wt\psi_2 &  \dots & 0 & 0 \\
\vdots & \vdots & \ddots & \vdots & \vdots \\
0 & 0 &  \dots & \wt\psi_{n-1} & 0 \\
\end{pmatrix}\,.
\end{equation}
If $\theta$ is arbitrary, we must have in addition that $\psi_n=\wt\psi_n=0$.  With this reduction to our Hitchin equations action  \eqref{eq:action for Hitchin equations}, the WZ term vanishes because $h$ is diagonal, and it reduces to
\begin{equation} \label{eq:Toda-action}
    \im\bigintsss_\Sigma\sum_{i=1}^n\Big(\p v_i\wedge\dbar v_i - 2e^{v_{i+1} - v_i}\p\psi_i\wedge\dbar\wt\psi_i\Big)\,. 
\end{equation}
Slightly abusing notation by writing $\diag\{\e^{g_{0i}}\}$ and $\diag\{\e^{g_{\infty i}}\}$ for $g_0$ and $g_\infty$ respectively, the residual gauge freedoms \eqref{psi-gauge} and \eqref{h-gauge} become
\begin{equation} \begin{aligned}
    &v_i \mapsto v_i + g_{0i}(w) - g_{\infty i}(\bar w)\,,\qquad \sum_ig_{0i} = \sum_ig_{\infty i} = 0\,, \\
    &(\wt\psi_i,\psi_i) \mapsto \big(e^{g_{0i}(w) - g_{0(i+1)}(w)}\wt\psi_i, e^{g_{\infty(i+1)}(\bar w) - g_{\infty i}(\bar w)}\psi_i\big) + (\chi_i(w),\wt\chi_i(\bar w))\,. \label{Toda-gauge}
\end{aligned} \end{equation}
The equations of motion for $\psi_i,\wt\psi_i$ read
\be \label{eq:chi-eoms} \dbar(e^{v_{i+1} - v_i}\p\psi_i) = \p(e^{v_{i+1} - v_i}\dbar\wt\psi_i) = 0\,. \ee
Working on $\Sigma = \R^2$, we can solve these to get
\be \p_w\psi_i = \frac{1}{2}F_i(w)e^{v_i - v_{i+1}}\,,\qquad \p_{\bar w}\widetilde\psi_i = \frac{1}{2}\wt F_i(\bar w)e^{v_i-v_{i+1}}\,. \ee
The $(g_i,\wt g_i)$ can be chosen to make all the $F_i=F$ and $\wt F_i=\wt F$ for some $F(w), \wt F (\bar w)$ in the affine case, and in the finite case we have enough freedom to reduce to $F=\wt F=1$.  We can now plug these back into the $v_i$ equation of motion to get
\begin{equation}
\Delta v_i = F(w) \wt F(\bar w)(e^{v_i - v_{i+1}} - e^{v_{i-1} - v_i})\,.
\end{equation}
In the finite case with $F = \wt F = 1$ these are the Toda field equations.  For the affine Kac-Moody algebra $\wh{\mathfrak{su}}(n)$ they are a (subset of) the equations of motion of conformal affine Toda theory (CAT) \cite{BabelonBonora199007,AratynFerreiraGomesZimerman199107,Constantinidis:1992hs}.\footnote{The equations of motion of CAT take the form $\Delta t_i = e^\eta e^{\sum_jK_{ij}t_j}$ for $K$ the Cartan matrix of $\wh{\mathfrak{su}}(n)$ and $\eta$ an additional field obeying $\Delta\eta = 0$.  We can identify $F(w)\wt F(\bar w) = e^\eta$.  The constraint $\sum_it_i = 0$ is also not imposed, but since the interactions depend only on differences, this extra centre of mass variable does not enter the $v_i = t_i - t_{i-1}$ equations of motion.  We are grateful to Luiz Agostinho Ferreira for alerting us to the connection to CAT.}  Making the coordinate choice
\begin{equation}
    \rd w'= F(w)\rd w \label{coord-fix}
\end{equation}
recovers the standard field equations of affine Toda theory.  This dichotomy between finite and affine Toda reflects the conformal invariance of the former in that the conformal transformation $w\rightarrow w'$ is naturally absorbed into the gauge freedom \eqref{Toda-gauge} whereas in the latter a coordinate is naturally singled out by \eqref{coord-fix}.

\begin{rmk}
The action \eqref{eq:Toda-action} does not match the familiar expressions given in equation \eqref{eq:Toda-Lagrangian}.  Indeed, it depends on an extra field $\psi,\wt\psi$ which can be eliminated on-shell.  These extra degrees of freedom play a crucial role in the Alday-Maldacena formulation of minimal surfaces, but it would nevertheless be desirable to recover the Lagrangians \eqref{eq:Toda-Lagrangian} directly.  We expect that in order to do so, conformal symmetry must be explicitly broken through the boundary conditions. \qed 
\end{rmk}

\section{Discussion and Future Directions}
\label{sec:discussion}

In this work, we first constructed an action for Hitchin's equations in \S\ref{sec:2d action for Hitchin's equations}. We then derived this action from 4d Chern--Simons theory in \S\ref{sec:4dCS-Hitchin}. The symplectic structure arising from this four-dimensional Chern--Simons theory in \S\ref{sec:symplectic structure, and Hitchin Hamiltonians}  coincides with the symplectic form on the space of solutions of Hitchin's equations that is the K\"ahler form on the realization of $\hit$ as a moduli space of Higgs bundles. We further constructed the Hitchin Hamiltonians in terms of the four-dimensional gauge field. To realize the other  $\CP^1$-family of symplectic structures, we constructed corresponding four-dimensional Chern--Simons theory, each corresponding to its own family of 2d actions, with Hitchin's equations as their equations of motion, giving the whole hyperk\"ahler family of symplectic structures on Hitchin moduli space. These are parametrized by  $\CP^1$, which provides the base of the holomorphic fibration of the twistor space of the Hitchin moduli space.  This $\CP^1$ base is thereby identified with the holomorphic plane of four-dimensional Chern--Simons theory on which the spectral parameter lives.

\smallskip A similar identification between the $\CP^1$ base of the twistor space of the Hitchin moduli space and the spectral parameter Riemann sphere was made in   \cite{FrostGurdoganMason202306}.  There, the case of the Hitchin system arising in the Alday--Maldacena formulation of minimal surfaces in AdS was studied \cite{AldayMaldacena200904,AldayGaiottoMaldacena200901}.
These, in fact, correspond to $\SU(2)$ Hitchin systems with $\Z_2$ symmetry for minimal surfaces in AdS$_3$ and $\SU(4)$ with $\Z_4$ symmetry for minimal surfaces in AdS as examples of our affine Toda treatment in \S\ref{sec:affineToda}.
In that problem, the moduli space could be identified with the boundary conditions of the minimal surfaces and needs a nontrivial function $F(w)$ to appear dynamically. This work formed part of the motivation for this work, so as to find action formulations that interpolate between that of the minimal surface and the twistor description of the moduli space.  Eventually, one might be able to use such a description of the twistor space for the moduli space as a route to the quantization of such strings in AdS. These might then have application to the study of Wilson loops and amplitudes at finite coupling.

\smallskip Another natural extension will be to consider Hitchin’s equations on $\Sigma$ in the presence of prescribed singularities in their solutions at a finite set of points. This is the subject of the upcoming work \cite{Moosavian2026}. Indeed, the case discussed above in \cite{FrostGurdoganMason202306} is a rank two example where $\Sigma$ is a Riemann sphere with a single wildly ramified puncture. There are further examples in this vein with higher rank, and an extra normal puncture with applications to form factors at strong coupling that would be interesting to understand better.

\smallskip Much of the theory of integrable systems remains to be formulated in this 4d framework.  The key 2d models of \cite[\S 6]{MasonWoodhouse199605}, such as the Korteweg de Vries and Nonlinear Schr\"odinger equations have yet to be given an efficient  4d Chern--Simons formulation. Their hierarchies, and those of other models, are commuting flows that arise from the sequences of conserved quantities.  These fit naturally into the self-duality and twistor framework, for example, see the descriptions in of \cite[\S 8]{MasonWoodhouse199605}.  It seems likely that in 4d Chern--Simons theory they should arise by using the freedom of the boundary conditions  in \eqref{bck}: given a complex structure on $\Sigma$ with local holomorphic coordinate $w$, there is a natural hierarchy of choices of boundary condition when  $\omega$ has a pole of order $q$, at $\spa=0$,
 \begin{equation}
     \A_w \sim \mcal{O}(\spa^{-p})\,,
\qquad \A_{\bar w}  \sim \mcal{O}(\spa^{p+q})\, , \qquad p\in \Z %
\, ,\label{bck1}
\end{equation}
These are then sufficient to make the action bounded and gauge invariant near the pole at $\spa=0$ and have the freedom to lead to a hierarchy of models.  

\smallskip The reductions to one dimension are also interesting, as discussed in \cite[\S 7]{MasonWoodhouse199605}, including Nahm's equations, spinning tops, and Painlev\'e equations.

\smallskip Ultimately, the most interesting applications of these ideas will be to the quantization of these systems, both directly and via a quantization of the twistor formulation of their moduli spaces. 
\\

\paragraph{\large Acknowledgement.}  We would like to thank Kevin Costello and David Gaiotto for comments on a draft.  Furthermore, we are grateful to Vincent Caudrelier, Derek Harland, and Benoît Vicedo for many useful comments and helpful clarification regarding Remark \ref{rmk:other-actions} which appeared in an earlier version of this work.  Research at Perimeter Institute is supported by the Government of Canada through Industry Canada and by the Province of Ontario through the Ministry of Research and Innovation.  LJM would like to thank the Simons Collaboration on Celestial Holography CH-00001550-11 and STFC for financial support from grant number ST/T000864/1.  The research of SFM is funded by the ERC Consolidator Grant \#864828 ``Algebraic Foundations of Supersymmetric Quantum Field Theory'' (SCFTAlg).

\bibliography{References}
\bibliographystyle{JHEP}

\end{document}